\documentclass[prd,11pt]{revtex4}
\usepackage{amssymb,amsmath,amsthm,graphicx,psfrag}

\pdfoutput=1

\begin{document}


\def\be{\begin{equation}}
\def\ee{\end{equation}}
\def\bea{\begin{eqnarray}}
\def\eea{\end{eqnarray}}
\newcommand{\der}[2]{\frac{\partial{#1}}{\partial{#2}}}
\newcommand{\dder}[2]{\partial{}^2 #1 \over {\partial{#2}}^2}
\newcommand{\dderf}[3]{\partial{}^2 #1 \over {\partial{#2} \partial{#3}}}
\newcommand{\eq}[1]{Eq.~(\ref{eq:#1})}
\newcommand{\dd}{\mathrm{d}}

\title{Thermodynamic instability of rotating black holes}

\vskip 4cm

\author{R. Monteiro}
\email{R.J.F.Monteiro@damtp.cam.ac.uk}
\author{M. J. Perry}
\email{M.J.Perry@damtp.cam.ac.uk}
\author{J. E. Santos}
\email{J.E.Santos@damtp.cam.ac.uk}

\vskip 0.2cm

\affiliation{DAMTP, Centre for Mathematical Sciences, University of Cambridge, Wilberforce Road, Cambridge CB3 0WA, UK}

\vskip 0.5cm

\date{\today}

\vskip 4cm

\begin{abstract}
We show that the quasi-Euclidean sections of various rotating black holes in different dimensions possess at least one non-conformal negative mode when thermodynamic instabilities are expected. The boundary conditions of fixed induced metric correspond to the partition function of the grand-canonical ensemble. Indeed, in the asymptotically flat cases, we find that a negative mode persists even if the specific heat at constant angular momenta is positive, since the stability in this ensemble also requires the positivity of the isothermal moment of inertia. We focus in particular on Kerr black holes, on Myers-Perry black holes in five and six dimensions, and on the Emparan-Reall black ring solution. We go on further to consider the richer case of the asymptotically AdS Kerr black hole in four dimensions, where thermodynamic stability is expected for a large enough cosmological constant. The results are consistent with previous findings in the non-rotation limit and support the use of quasi-Euclidean instantons to construct gravitational partition functions.

\vskip 0.5cm
 
\end{abstract}

\maketitle


\section{Introduction}

Gravitation is a purely attractive force, which has led to a number of conundrums revolving around questions of stability. In classical physics, it seems that the question is settled. Gravitational collapse cannot, under a wide range of circumstances, be prevented. Although collapse to form a singularity happens, it is believed that these singularities will be isolated from observation by horizons. Thus the end-point of gravitational collapse is believed to always result in black holes. Classically, many black hole solutions are stable, in particular the ones of astrophysical relevance \cite{PhysRev.108.1063,PhysRevLett.24.737,PhysRevD.1.2870,Teukolsky:1972my,Chandrasekhar:1985kt,Whiting:1988vc,Ishibashi:2003ap}.

Quantum mechanics changes this. In 1974, Hawking discovered that black holes have a temperature, the Hawking temperature $T_H$, given by
\be
T_H=\frac{\kappa}{2\pi}, \nonumber
\ee
where $\kappa$ is the surface gravity of the black hole \cite{Hawking:1974rv} (we use natural units, so that $G=c=\hbar=k_B=1$ throughout). For non-rotating black holes, this gives rise to a new instability. Since
\be
\kappa = \frac{1}{4 M}, \nonumber
\ee
where $M$ is the black hole mass, isolated black holes will radiate and lose energy. This will cause them to heat up. Since conservation of energy leads to
\be
\dot{M}\sim-\frac{1}{M^2} \nonumber
\ee
we see this is a runaway process. When the black hole reaches zero mass, it is presumed to disappear completely. The specific heat of the black hole is negative
\be
C=-\frac{1}{8\pi M^2}, \nonumber
\ee
a typical sign of instability \cite{Hawking:1976de}.

It is a sign that the canonical ensemble breaks down for such objects leading to doubts as to whether a conventional thermodynamic interpretation is possible. However, if instead one looks at the microcanonical ensemble, one discovers that it is well-defined.

If one includes the possibility that the black holes are rotating with angular momentum $J$ or have an electric charge $Q$, one finds that if
\be
J^4+6 J^2 M^4+4 Q^2 M^6-3 M^8> 0 \nonumber
\ee
then the specific heat at constant $J$ and $Q$ turns out to be positive.

One expects that these difficulties will be reflected in the path-integral treatment of gravitation. Suppose one tries to calculate the canonical, or grand-canonical, partition function. Then one needs to integrate over all physical fields subject to certain boundary conditions. Generally for the grand-canonical ensemble, one integrates over quasi-Euclidean configurations (see section \ref{sec:compB} for a more complete description of what this means). So, the field configurations must be periodic in imaginary time, with periodicity equal to the inverse temperature, and quasi-periodic in the complexified azimuthal angle generated by any conserved angular momenta, or quasi-periodic under complexified gauge transformations associated to any conserved charge.

The gravitational path integral based on the Einstein action is not well-defined because of its lack of renormalizability. However, at the semi-classical level, it makes sense as an effective field theory, perhaps derived from some more fundamental theory such as string theory. This path integral has, at first sight, a big difficulty with stability, as the kinetic energy operator for conformal transformations has the wrong sign. However, it turns out that fluctuations in the path integral of such a type are gauge artifacts. Much more serious and interesting is the possibility that the gauge invariant parts of the fluctuations contribute with the `wrong' sign to the partition function. This is a sign of instability. For the four-dimensional non-rotating black holes, these negative modes have been known for some time \cite{Gross:1982cv,Prestidge:1999uq}.

In this paper, we extend our knowledge of this type of instability. In Section~\ref{sec:comp}, we describe the formalism required to identify the gauge invariant negative modes. In Section~\ref{sec:probe}, we describe a way of finding gauge invariant deformations of a specified field configuration. These deformations are not themselves the negative mode, but since they decrease the Euclidean action, they prove that negative modes exist. In Section~\ref{sec:cases}, we apply our technique to the four-dimensional Kerr solution, five and six-dimensional Myers-Perry metrics, the singly-spinning five-dimensional black ring and the four-dimensional Kerr-AdS solution. In every case, we find negative modes, except for large Kerr-AdS black holes. In Section~\ref{sec:interpretation}, we look at the thermodynamics of the black holes or rings, and see how it matches up with the existence of our negative modes. We use this to make speculations about when the thermodynamic approximation is, or is not, valid.

\section{\label{sec:comp}The gravitational path integral}

\subsection{\label{sec:compA}The decomposition theorem}

The path integral of Euclidean quantum gravity,
\be
\label{pathintegral}
Z = \int \mathrm{D}[g] e^{-I[g]},
\ee
is constructed from the action
\be
\label{action}
I[g] = -\frac{1}{16 \pi} \int_{\mathcal M} \dd^d x \sqrt{g}\, (R-2 \Lambda) -\frac{1}{8 \pi} \int_{\partial {\mathcal M}} \dd^{d-1} x \sqrt{g^{(d-1)}}\, K - I_0.
\ee
The first term is the usual Einstein-Hilbert action and the second is the York-Gibbons-Hawking boundary term \cite{York:1972sj,Gibbons:1976ue}, where $K$ is the trace of the extrinsic curvature on $\partial \mathcal M$. This term is required for non-compact manifolds $\mathcal M$, as the ones we will study, so that the boundary condition on $\partial \mathcal M$ is a fixed induced metric, and not fixed derivatives of the metric normal to $\partial \mathcal M$.

The term $I_0$ can depend only on $g^{(d-1)}_{ab}$, the induced metric on $\partial \mathcal M$, and not on the bulk metric $g_{ab}$, so that it can be absorbed into the measure of the path integral. However, since we are interested in the partition functions of black holes, it is convenient to choose it so that $I=0$ for the background spacetime that the black hole solution approaches asymptotically. For asymptotically flat black holes \cite{Gibbons:1976ue}, the Einstein-Hilbert term is zero and the action becomes
\be
-{1 \over 8 \pi} \int_{\partial {\mathcal M}} \dd^{d-1} x \sqrt{g^{(d-1)}}\, ( K - K_0 ),
\ee
where $K_0$ is the trace of the extrinsic curvature of the flat spacetime matching the black hole metric on the boundary $\partial \mathcal M$ at infinity. This subtraction renders the action of the black hole finite. For asymptotically AdS black holes \cite{Hawking:1982dh,Gibbons:2004ai}, the boundary terms cancel when the background subtraction is performed, but the bulk volume integral diverges and requires an analogous subtraction that sets the action of AdS space to zero. An alternative view is that $I_0$ should be seen as a counterterm, corresponding to the counterterm of a dual conformal field theory (see \cite{Balasubramanian:1999re,Kraus:1999di,Skenderis:2000in,Olea:2005gb,Olea:2006vd}).

The gravitational path integral (\ref{pathintegral}) is non-renormalisable but we expect meaningful results in an effective field theory approach. A different issue is that the action (\ref{action}) can be made arbitrarily negative so that the path integral appears to be always divergent even at tree-level. These problems can be addressed in the semiclassical approximation, where the path integral is dealt with by saddle-point methods. We consider a saddle-point $\widehat{g}_{ab}$, \emph{i.e.} a non-singular solution of the equations of motion,
\be
\label{rab}
\widehat{R}_{ab}= \frac{2 \Lambda}{d-2} \widehat{g}_{ab},
\ee
usually referred to as a gravitational instanton. We then treat as a quantum field $h_{ab}$ the small perturbations about the saddle-point,
\be
g_{ab} = \widehat{g}_{ab} + h_{ab}.
\ee 
This leads to a perturbative expansion of the action,
\be
I[g]= I[\widehat{g}] + I_2[h;\widehat{g}] + {\mathcal O}(h^3).
\ee
The first order action $I_1$ vanishes since $\widehat{g}_{ab}$ obeys the equations of motion, while the second order action $I_2$, which gives the one-loop correction, is the action for the quantum field $h_{ab}$ on the background geometry $\widehat{g}_{ab}$.

The effective field theory is valid if the background geometry $\widehat{g}_{ab}$ has a curvature nowhere near the Planck scale. We can also address the issue of the arbitrarily negative action geometries in the path integral, called `conformal factor problem' since it is the conformal direction in the space of metrics that is responsible for the divergence. Perturbatively, this corresponds to trace-like perturbations $h_{ab}$ which lead to a negative $I_2$. The prescription of \cite{Gibbons:1978ac} is that the integration contour for those perturbations is imaginary. They can then be seen to be irrelevant and don't represent physical instabilities. Of physical interest are the instabilities studied firstly in \cite{Gross:1982cv}, the analysis of which we wish to extend to rotating black holes. 

We follow here the procedure in \cite{Gibbons:1978ji}, straightforwardly extended to higher dimensions. We will decompose the second order action, applying a standard gauge fixing procedure, and show that the unphysical divergent modes do not contribute to the one-loop partition function.

The partition function is
\be
Z_\mathrm{ 1-loop} = e^{-I[\widehat{g}]} \int \mathrm{D}[h] (\mathrm{G.F.})  e^{-I_2[h;\widehat{g}]},
\ee
where $(\mathrm{G.F.})$ denotes all contributions induced by fixing the gauge in the path integral. Hereafter, $\widehat{g}_{ab}$ is relabelled as $g_{ab}$ and all metric operations are performed with it. The second order action is given by
\be
\label{actioninitial}
I_2[h;g] = - \frac{1}{16 \pi} \int \dd^d x \sqrt{g}\, \left[ - \frac{1}{4} \bar{h} \cdot Gh + \frac{1}{2} (\delta \bar{h})^2  \right],
\ee
where $\cdot$ denotes the metric contraction of tensors. We have defined 
\be
\bar{h}_{ab} = h_{ab} - \frac{1}{2} g_{ab} h^c_{\phantom{i}c}
\ee
and
\be
(G h)_{ab} = - \nabla^c \nabla_c h_{ab} -2 R_{a\phantom{c}b}^{\phantom{a}c\phantom{b}d} h_{cd},
\ee
where the operator $G$ is related to the Lichnerowicz Laplacian $\Delta_L$ by $G= \Delta_L - 4 \Lambda / (d-2)$. We also define the operations on tensors $T$
\begin{subequations}
\be
(\delta T)_{b \dots c} = - \nabla^a T_{ab \dots c},
\ee
\be
(\alpha T)_{ab \dots c} = \nabla_{(a} T_{b \dots c)}.
\ee
\end{subequations}

The second order action $I_2[h;g]$ is invariant for the diffeomorphism transformations
\be
h_{ab} \to h_{ab} + \nabla_{a} V_{b} + \nabla_{b} V_{a}  = (h + 2 \alpha V)_{ab}.
\ee
Following the Feynman-DeWitt-Faddeev-Popov gauge fixing method,
\be
\label{fpgf}
(\mathrm{G.F.}) = (\mathrm{det}\, C) \,\delta(C_a[h] - w_a).
\ee
We consider the linear class of gauges
\be
\label{wa}
C_b[h]= \nabla^a \left( h_{ab}- {1 \over \beta} g_{ab} h^c_{\phantom{i}c} \right),
\ee
where $\beta$ is an arbitrary constant, so that the Fadeev-Popov determinant $(\mathrm{det}\, C)$ is given by the spectrum of the operator
\be
(C V)_a = - \nabla^b \nabla_b V_a - R_{ab} V^b +\left( \frac{2}{\beta} -1 \right) \nabla_a \nabla_b V^b.
\ee
To study the spectrum, let us consider the Hodge-de Rham decomposition of the gauge vector $V$ into harmonic (H), exact (E) and coexact (C) parts,
\be
V = V_\mathrm{H} + V_\mathrm{E} + V_\mathrm{C}.
\ee
This induces a decomposition of the action of $C$, which we denote by $C_\mathrm{H}$ for harmonic vectors, $C_\mathrm{E}$ for exact vectors and $C_\mathrm{C}$ for coexact vectors.

The harmonic part satisfies $\dd V_\mathrm{H} = 0$ and $\delta V_\mathrm{H} = 0$. We can check that
\be
\label{spch}
C V_\mathrm{H} = -\frac{4 \Lambda}{d-2} V_\mathrm{H}.
\ee
The spectrum is positive for $\Lambda<0$ and zero for $\Lambda=0$, with multiplicity given by the number of linearly independent harmonic vector fields. For $\Lambda > 0$, the background solution satisfying (\ref{rab}) does not allow for harmonic vector fields if assumed to be compact and orientable \cite{Yano}. Thus, the spectrum of $C_\mathrm{H}$ is never negative.

The exact part is such that $V_\mathrm{E} = \dd \chi$, where $\chi$ is a scalar. We can show that
\be
\label{Ceop}
\mathrm{spec}\;C_\mathrm{E} = \mathrm{spec}\; \left( 2 \left[ \left(\frac{1}{\beta} -1 \right) \Box -\frac{2 \Lambda}{d-2} \right] \right),
\ee
where the operator on the RHS acts on scalars, and $\Box$ is the Laplacian. For $\Lambda < 0$, the operator is positive for $\beta>1$, being positive semi-definite for $\Lambda=0$. For $\Lambda > 0$, the Lichnerowicz-Obata theorem tells us that the spectrum of the Laplacian on a compact and orientable manifold satisfying (\ref{rab}) is bounded from above by $- 2d \Lambda/((d-1)(d-2))$, the saturation of the bound corresponding to the sphere \cite{Yano}. This implies that, for $\Lambda > 0$, the spectrum of $C_\mathrm{E}$ is positive for $\beta > d$.

The coexact part is such that $\delta V_\mathrm{C} = 0$. Hence
\be
\label{spcc}
C V_\mathrm{C} = 2 \delta \alpha V_\mathrm{C}
\ee
and the spectrum of $C_\mathrm{C}$ can be shown to be positive semi-definite,
\be
\label{ineqCc}
\int \dd^d x \sqrt{g}\, [ V_\mathrm{C} \cdot C V_\mathrm{C} ] = 2 \int \dd^d x \sqrt{g}\, [ \alpha V_\mathrm{C} \cdot \alpha V_\mathrm{C} ] \geq 0,
\ee
with equality for coexact Killing vectors.

The Faddeev-Popov determinant contribution to the partition function is then
\be
\label{detC0}
\mathrm{det}\, \tilde{C} \sim (\mathrm{det}\, \tilde{C}_\mathrm{E}) (\mathrm{det}\, \tilde{C}_\mathrm{C}),
\ee
the tilde denoting that the zero modes have been projected out. The harmonic contribution is not explicitly considered because, if it exists ($\Lambda<0$), it is a positive factor dependent only on $\Lambda$ and on the dimension of the space of harmonic vector fields, as mentioned above; it will not be relevant to our discussion. The contribution from the exact part is fundamental since it will cancel the divergent modes of the field $h_{ab}$.

In order to make the results independent of the arbitrary vector $w$ in the gauge fixing (\ref{fpgf}), the 't Hooft method of averaging over gauges is adopted. The arbitrariness is then expressed in terms of a constant $\gamma$ introduced by the weighting factor of the averaging. The final result will be independent of $\gamma$, as required. The unconstrained effective action for the perturbations is given by
\begin{align}
& I_2^\mathrm{eff}[h;g] = I_2[h;g] + \frac{\gamma}{32 \pi}  \int \dd^d x \sqrt{g}\, C^a[h] C_a[h] = \nonumber \\
= & - \frac{1}{16 \pi} \int \dd^d x \sqrt{g}\, \left[ - \frac{1}{4} \bar{h} \cdot Gh + \frac{1}{2} (1-\gamma) (\delta \bar{h})^2  +\frac{\gamma}{2} \left( 1- \frac{2}{\beta} \right) \delta \bar{h} \cdot \dd \hat{h} - \frac{\gamma}{8} \left( 1- \frac{2}{\beta} \right)^2 (\dd \hat{h})^2  \right],
\end{align}
where we denote $\hat{h} \equiv h^c_{\phantom{i}c}$.

We now decompose the quantum field $h_{ab}$ into a traceless-transverse (TT) part, a traceless-longitudinal (TL) part, built from a vector $\eta$, and a trace part,
\be
h_{ab} = h_{ab}^{TT} + h_{ab}^{TL} + {1 \over d} g_{ab} \hat{h},
\ee
with
\be
h_{ab}^{TL} = 2 (\alpha \eta)_{ab} + {2 \over d} g_{ab} \delta \eta.
\ee

The constant $\beta$, unspecified in the gauge condition (\ref{wa}), can be chosen so that the trace $\hat{h}$ and the longitudinal vector $\eta$ decouple. This requires
\be
\label{beta}
\beta = 2 \left( 1- {d-2 \over d} { \gamma-1 \over \gamma } \right)^{-1}.
\ee
The effective action becomes
\begin{align}
I_2^\mathrm{eff}[h;g] & = - \frac{1}{16 \pi} \int \dd^d x \sqrt{g}\, \Bigg[ - \frac{1}{4} h^{TT} \cdot G h^{TT} - \alpha \eta \cdot \alpha \Delta_1 \eta - \frac{1}{d} \delta \eta \Box \delta \eta +
\nonumber \\
& + 2 (1-\gamma) \left( \delta \alpha \eta \cdot \delta \alpha \eta + \frac{1}{d^2} \alpha \delta \eta \cdot \alpha \delta \eta - \frac{2}{d} \delta \alpha \eta \cdot \alpha \delta \eta \right) +
\nonumber \\
& +\frac{4}{d-2} \Lambda \left( \alpha \eta \cdot \alpha \eta  - \frac{1}{d} (\delta \eta)^2 \right) +\frac{1}{2} \hat{h} F \hat{h}  \Bigg],
\end{align}
where the operator $F$ is given by
\be
F = - {d-2 \over 4d} \left( 1+ {d-2 \over d} {\gamma-1 \over \gamma} \right) \Box - {1\over d} \Lambda.
\ee
Recalling the choice of $\beta$ (\ref{beta}), we find that the operator on the RHS of the expression (\ref{Ceop}) is given by $4d F /(d-2)$. The contribution of the ghosts (\ref{detC0}) can be recast as
\be
\label{detC1}
\mathrm{det}\, \tilde{C} \sim (\mathrm{det}\, \tilde{F}) (\mathrm{det}\, \tilde{C}_\mathrm{C}).
\ee

For the vector $\eta$, as we did for $V$ in the ghost part, we perform a Hodge-de Rham decomposition into harmonic, coexact and exact parts,
\be
\eta = \eta_\mathrm{H} + \eta_\mathrm{C} + \eta_\mathrm{E},
\ee
respectively. Using $\eta_\mathrm{E} = \dd \chi$, the result for the effective action is then
\begin{align}
\label{I2eff}
I_2^{\mathrm{eff}}[h;g] = & -{1 \over 16\pi} \int \dd^d x \sqrt{g}\, \Big[ - {1\over 4} h^{TT} \cdot G h^{TT} + {1\over 2} \hat{h}F\hat{h} + \nonumber \\ & + {4\over d-2} \gamma \Lambda \alpha \eta_\mathrm{H} \cdot \alpha \eta_\mathrm{H} - \gamma \alpha \eta_\mathrm{C} \cdot \alpha C_\mathrm{C} \eta_\mathrm{C} - {4d\over d-2} \gamma D\chi \cdot DF\chi \Big],
\end{align}
where we defined the operator
\be
D_{ab} = \nabla_a \nabla_b - {1 \over d} g_{ab} \Box.
\ee
Notice that the Hodge-de Rham decomposition of $\eta$ in harmonic, coexact and exact parts gives, for $h^{TL}_{ab}$, a decomposition in $2\alpha \eta_\mathrm{H}$, $2\alpha \eta_\mathrm{C}$ and $2D\chi$, respectively.

Finally, we can evaluate the Gaussian integrals in the partition function to show the dependence
\begin{align}
\label{Zfinal}
Z_\mathrm{ 1-loop} & \sim (\mathrm{det}\, \tilde{C}) (\mathrm{det}\,\tilde{G})^{-1/2}  (\mathrm{det}\, \tilde{F})^{-1/2} (\mathrm{det}\, \tilde{C}_\mathrm{C})^{-1/2} (\mathrm{det}\,\tilde{F})^{-1/2} 
\nonumber \\
& \sim (\mathrm{det}\,\tilde{G})^{-1/2} (\mathrm{det}\, \tilde{C}_\mathrm{C})^{1/2}.
\end{align}
Again, the tilde on the operators denotes that the zero modes have been projected out. The Gaussian integrals are regularised by $\zeta$-function methods \cite{Gibbons:1978ji}. It is understood that the spectrum of $G$ here is restricted to traceless-transverse normalisable modes.

Let us review the treatment of the `conformal factor problem'. Trace-type perturbations make the action (\ref{actioninitial}) negative. But a detailed analysis showed that these modes do not contribute to the path integral. The two factors $(\mathrm{Det} F)^{-1/2}$ arising from the Gaussian integrals in $\hat{h}$ and $\chi$ cancel with the $\mathrm{Det} F$ factor arising from the exact part of the Fadeev-Popov determinant. This makes the unphysical character of the divergence obvious, at least in perturbation theory. The conclusion is that the non-positivity of the action (\ref{action}) and the resulting apparent divergence of the gravitational path integral are fixed by projecting out this contribution.

The relevant operators are then $C_\mathrm{C}$ and $G$. For a real metric, the operator $C_\mathrm{C}$ is positive semi-definite, as we have shown above. Once its zero modes are projected out, it contributes a positive factor to the final result. The physical instabilities, identified by imaginary contributions to the partition function, are only possible if there are negative eigenvalues of the operator $G$, $G h^{TT}= \lambda h^{TT}$. This was the problem studied in \cite{Gross:1982cv} for the Schwarzschild black hole. We intend to extend this treatment to rotating black holes, which requires addressing the problem of complex instantons.

\subsection{\label{sec:compB}Quasi-Euclidean geometries}

The partition function is usually defined as a Euclidean path integral, a sum over real geometries for which imaginary time $\tau=it$ is used. However, while static geometries remain real for this analytical continuation, the same does not hold for stationary geometries. In the canonical formalism, where $\gamma_{ij}$ is the metric on a constant time slice, $N$ is the lapse function and $N^i$ is the shift vector required for rotating spacetimes, we have
\be
\label{ds2i}
\dd s^2 = N^2 \dd \tau^2 + \gamma_{ij}(\dd x^i-iN^i \dd \tau)(\dd x^j-iN^j \dd \tau).
\ee
These geometries have been called \emph{quasi-Euclidean}. The question is then whether one should analytically continue the shift vector (e.g. through the rotation parameters for a Kerr black hole) in order to get a real geometry. This is trivial when one considers the instanton approximation to the path integral, because the parameters made imaginary can simply be made real again in the final result. But when one goes beyond leading order, as is the case in this paper, and considers metrics that do not satisfy the equations of motion but are also included in the sum, the choice affects the positivity properties of the second order action and thus the convergence of the path integral.

We share the view of \cite{Brown:1990di} and \cite{Brown:1990fk} that the continuations other than the usual $\tau=it$ lead to unphysical parameters. As those authors point out, the leading order instanton action is real in spite of being constructed with a complex metric, and it corresponds to the physical free energy. The charges and the horizon locus remain the same as in the Lorentzian case. Studying the convergence of the path integral for particular imaginary values of the Kerr rotation parameters, for instance, bears no relation to the actual black holes. A further argument can be made based on the black ring case. As opposed to the Kerr geometry, the black ring does not possess a real section with imaginary time that is regular, since conical singularities cannot be removed \cite{Astefanesei:2005ad,bella}.

The results in the previous decomposition of the metric were obtained for real Euclidean metrics. However, the expression (\ref{Zfinal}) should still hold for an appropriate complex contour of integration. This contour is specified in a standard way by the steepest descent method. The relevant eigenvalues of $C_\mathrm{C}$ and $G$ are now determined with respect to a complex metric, \emph{i.e.} to the physical Lorentzian rotation parameters. The semi-positivity of $C_\mathrm{C}$ is no longer obvious and we have nothing more to say about it. Still, a negative eigenvalue of $G$ is sufficient to cause problems in the definition of the path integral and herald an instability.

\section{\label{sec:probe}The probe perturbation}

As was discussed in the previous section, the Euclidean path integral only depends on the spectrum of two operators: $C_\mathrm{C}$ and $G$. The latter acts on traceless-transverse (TT) perturbations of the metric and will be the object of our attention in this section. In the Schwarzschild case, mostly due to the spherical symmetry of the problem, it was possible to determine the negative mode by a straightforward method \cite{Gross:1982cv}. However, for solutions such as Kerr, Myers-Perry or the black ring, this seems challenging, due to the lack of symmetry of the background geometry.

The approach that we will adopt here is somehow different. To prove that $G$ possesses at least one negative mode we only need to show that a particular TT probe perturbation renders the operator negative. In order to visualise this more clearly, pick an arbitrary TT perturbation and decompose it in eigenmodes of $G$,
\be
\phi_{ab}= \sum_n a_n \phi^{(n)}_{ab}.
\label{eq:probe1}
\ee
We can now construct the Rayleigh-Ritz functional, given by
\be
\mathcal{I} = \frac{\int\dd^d x \sqrt{g} \, \phi^{ab} (G\phi)_{ab}}{\int \dd^d x \sqrt{g} \, \phi^{ce} \phi_{ce}} = \frac{\sum_n \lambda_n a_n^2}{\sum_p a_p^2}.
\label{eq:probe2}
\ee
If a perturbation $\phi_{ab}$ is found such that $\mathcal{I}$ is negative, then it must be the case that at least one of the $\lambda_n$ is negative. This reasoning can be used to prove that a given instanton has a negative mode, but cannot be used to prove the converse. In fact, if $\mathcal{I}$ is positive for a particular $\phi_{ab}$, it might be the case that the $a_n$ corresponding to the negative eigenmode is small, or even absent, in the expansion (\ref{eq:probe1}).

We also have to check that our particular perturbation lies along the path of steepest descent. Since we will consider only perturbations that preserve the symmetries of the background, it suffices to check that the components $(G\phi)_{ab}$ are real or imaginary if and only if the components $h^{ab}$ are, \emph{i.e.} if and only if the components of the metric $g_{ab}$ are. This will indeed be the case for our perturbations: the components $(\tau,x^i)$ are imaginary and the others are real. Then, each term of the sum $\phi^{ab} (G\phi)_{ab}$ is real and this particular direction in the space of perturbations keeps the phase of the integrand functional constant, since the second order action is real. This is the condition for the steepest descent path.

It is the objective of this section to construct a TT perturbation that will render $\mathcal{I}$ negative, starting with a Killing vector field $k_a$. We will focus on pure Einstein gravity, with $\Lambda=0$, in an arbitrary number of dimensions $d$, where the background field equations are
\be
R_{ab}=0.
\label{eq:probe3}
\ee

For such a class of spacetimes we can construct a probe Maxwell field satisfying Maxwell's equations. Since $k_a$ is a Killing vector, it obeys to
\be
\nabla_a k_b+\nabla_b k_a = 0,
\label{eq:probe5}
\ee
from which we construct the following two-form components
\be
F_{ab}=\nabla_a k_b-\nabla_b k_a = 2 \nabla_a k_b.
\label{eq:probe4}
\ee

This field strength $F_{ab}$ trivially obeys to the Bianchi identities and also satisfies Maxwell's equations, since
\be
\nabla_a F^{ab} = -2 R^{ba}k_a = 0.
\label{eq:probe6}
\ee
In four dimensions, there exists a TT tensor that can be associated with such a field strength, the electromagnetic energy-momentum tensor,
\be
T_{ab}=F_{a}^{\phantom{a}c}F_{bc}-\frac{1}{4}g_{ab}F^{pq}F_{pq}.
\label{eq:probe7}
\ee

The energy-momentum tensor defined in \eq{probe7} is transverse in any dimension, since that only depends on the Bianchi identities associated with $F_{ab}$ and on \eq{probe6}, but the traceless condition is only valid in four dimensions. The strategy is to add a transverse component to \eq{probe7}, which will remove the trace. This can be accomplished by introducing an auxiliary scalar field $\sigma$, and by defining
\be
\phi_{ab}=T_{ab}-(\nabla_a\nabla_b\sigma-g_{ab}\Box\sigma),
\label{eq:probe8}
\ee
so that
\be
\nabla^a\phi_{ab} = -R_{ba}\nabla^a\sigma = 0.
\label{eq:probe9}
\ee
Requiring the tracelessness of $\phi_{ab}$ in \eq{probe8} gives
\be
\Box \sigma = \frac{1}{4}\Big(\frac{d-4}{d-1}\Big)F^{pq}F_{pq} =\Big(\frac{d-4}{d-1}\Big) (\nabla^p k^q) (\nabla_p k_q),
\label{eq:probe10}
\ee
For an arbitrary $k_a$, \eq{probe10} seems hopeless to invert. However, because $k_a$ is a Killing vector, there is a simple particular solution,
\be
\sigma = \frac{1}{2}\Big(\frac{d-4}{d-1}\Big) k^a k_a.
\label{eq:probe12}
\ee

The strategy is now clear: given $k_a$ we can construct $F_{ab}$ and $\sigma$ by using Eqs.~(\ref{eq:probe4}) and (\ref{eq:probe12}), respectively. These quantities are the only ingredients in the construction of $\phi_{ab}$, see Eqs.~(\ref{eq:probe7}) and (\ref{eq:probe8}). We thus conclude that for each $k^a$ we can associate a TT probe perturbation, in an arbitrary number of dimensions.

We are only interested in perturbations that are normalisable, in the sense that \makebox{$\int\dd^d x \sqrt{g}\phi^{ab}\phi_{ab}<+\infty$}. The obvious candidates for the Killing vector fields in a black hole background are either the time translational Killing vector $\partial_\tau$, or the azimuthal Killing vector $\partial_\phi$, both guaranteed in the case of an arbitrary black hole solution in pure $d-$dimensional Einstein gravity \cite{Hawking:1972qk,Hawking:1973uf}. $\partial_\phi$ leads to a non-normalisable perturbation, whereas $\partial_\tau$ leads to normalisable perturbations. Let us prove the last statement. If the spacetime is asymptotically flat, then $g_{\tau\tau}\simeq 1+\mathcal{O}(1/r^{d-3})$ and $g_{\tau x^i} \simeq \mathcal{O}(1/r^{d-3})$, which means that $k_a-\delta_a^{\tau} \simeq \mathcal{O}(1/r^{d-3})$ and thus ${F^a}_b\simeq \mathcal{O}(1/r^{d-2})$. The last statement implies that $\sqrt{g}\phi^{ab}\phi_{ab}\simeq \mathcal{O}(1/r^{3 (d-2)})$, from which we can see that this mode is normalisable as long as $d\ge 3$.

We also generalised the construction above to include a cosmological constant. However, we were only able to check for the particular cases of $d=4,5\text{ and }6$ that the perturbation was TT, in the background of Kerr-AdS and Myers-Perry-AdS in the corresponding dimensions. Unfortunately, in $d=5$ the perturbation turns out to be non-normalisable, and $d=6$ is computationally too challenging, so that we will only focus on the Kerr-AdS case. The form of the TT perturbation is more involved,
\be
\phi_{ab}=\tilde{T}_{ab}-(\nabla_a \nabla_b\tilde{\sigma}-g_{ab}\Box\tilde{\sigma})-\frac{d-4}{4}\tilde{F}_{(a}^{\phantom{(a}c}\bar{F}_{b)c},
\ee
where
\begin{subequations}
\begin{equation}
\begin{array}{ccc}
\displaystyle{\tilde{T}_{ab}=\tilde{F}_a^{\phantom{a}c}\tilde{F}_{bc}-\frac{g_{ab}}{4}\tilde{F}^{pq}\tilde{F}_{pq}}, & & \displaystyle{\tilde{\sigma}=\frac{1}{2}\Big({d-4 \over d-1}\Big)\tilde{k}^ak_a},
\end{array}
\end{equation}
\begin{equation}
\begin{array}{ccc}
\bar{F}_{ab}=\nabla_a \bar{k}_b-\nabla_b \bar{k}_a, & & \tilde{F}_{ab}=\nabla_a \tilde{k}_b-\nabla_b \tilde{k}_a,
\end{array}
\end{equation}
and
\begin{equation}
\begin{array}{ccc}
\tilde{k}_a=k_a-\bar{k}_a, & &\bar{k}_a=\bar{g}_{ab}k^b.
\end{array}
\end{equation}
\end{subequations}
In the expressions above, $k_a$ is a Killing vector of the original metric $g_{ab}$ and $\bar{g}_{ab}$ is a reference metric, obtained from the original metric by setting the mass of the black hole to zero, that is, the AdS metric. Note that $\bar{k}_a$ is a Killing vector of $\bar{g}_{ab}$, but $\tilde{k}_a$ is not a Killing vector field of either metrics $g_{ab}$ or $\bar{g}_{ab}$. Moreover, $\tilde{F}_{ab}$ satisfies the sourceless Maxwell's equations. The authors strongly believe that the perturbation described above should be TT for all Kerr-Schild spacetimes in any dimension, whose reference metric $\bar{g}_{ab}$ is that of a maximally symmetric spacetime.

In the next Sections, we will be able to identify negative modes using this probe perturbation.

\section{\label{sec:cases}Negative modes of gravitational instantons}
We will now apply the method described in the previous section to the Kerr black hole \cite{Kerr:1963ud}, the Myers-Perry black hole in five and six dimensions \cite{Myers:1986un}, and the five dimensional black ring of Emparan and Reall \cite{Emparan:2001wn}. All asymptotically flat black holes that we have studied have at least one normalisable negative mode, which suggests it may be a universal feature. In the last subsection, we study the Kerr-AdS black hole.


\subsection{\label{subsec:kerr}Kerr black hole}

The complexified version of the Kerr metric is given by
\be
\dd s^2 = \frac{\Sigma^2 \Delta \dd \tau^2}{\rho^2}+\frac{\rho^2\sin^2\theta}{\Sigma^2}\Big[\dd \phi+\frac{i a (r^2+a^2-\Delta)\dd \tau}{\rho^2}\Big]^2+\frac{\Sigma^2}{\Delta}\dd r^2+\Sigma^2 \dd\theta^2,
\ee
where
\begin{subequations}
\be
\Sigma^2 = r^2+a^2 \cos^2\theta,
\ee
\be
\Delta = r^2-r_0 r+a^2
\ee
and
\be
\rho^2 = (r^2+a^2)^2-\Delta a^2 \sin^2\theta.
\ee
\end{subequations}
The Kerr metric written in this way is already in the canonical (ADM) form. Here, $r_0$ is a mass scale, and is related to the black hole mass by $r_0 = 2 M$. Black holes require $a\le r_0/2$, where the inequality is saturated in the extremal limit. The complementary limit corresponds to naked singularities. The avoidance of a conical singularity at $r_{_+}=r_0/2+ \sqrt{(r_0^2/4)-a^2}$ requires the coordinate identification $(\tau,\phi)= (\tau+ \beta, \phi - i \beta \Omega)$, where $\beta=(r_{_+}^2+a^2)/ [2\pi(r_{_+}-r_0/2)]$ and $\Omega= a / (r_{_+}^2+a^2)$ are the black hole inverse temperature and angular velocity, respectively.

\begin{figure}
\centering
\includegraphics[width = 8 cm]{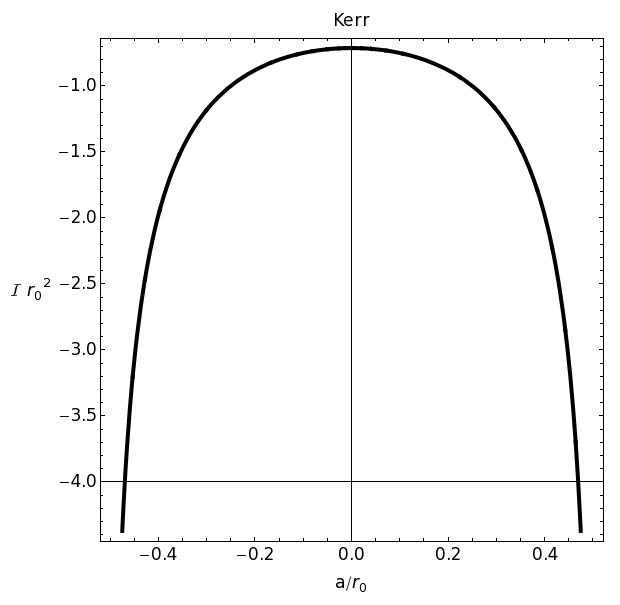}
\caption{\label{fig:kerr}For the Kerr instanton, $\mathcal{I}$ is negative, decreasing monotonically away from $a=0$ and evaluating to $-12$~$r_0^{-2}$ at extremality $|a| = r_0/2$.}
\end{figure}

For this particular case, due to its simplicity, we will present the explicit expression of the TT perturbation that we used to prove that $\mathcal{I}$ is negative,
\be
\phi^a_{\phantom{a}b}=\frac{r_0^2}{2 \Sigma^4}\left[
\begin{array}{cccc}
\displaystyle{1+\frac{2 a^2 \sin ^2\theta}{\Sigma^2}} & 0 & 0 & \displaystyle{-\frac{2 i a(r^2+a^2) \sin ^2\theta}{\Sigma^2}} \\
 0 & 1 & 0 & 0 \\
 0 & 0 & -1 & 0 \\
\displaystyle{-\frac{2 i a}{\Sigma^2}} & 0 & 0 & \displaystyle{-\Big(1+\frac{2 a^2 \sin ^2\theta}{\Sigma^2}\Big)}
\end{array}
\label{eq:kerrprobe}
\right].
\ee
This perturbation is clearly traceless and can be checked to be transverse and normalisable. The Rayleigh-Ritz functional defined in \eq{probe2}, evaluated for the perturbation (\ref{eq:kerrprobe}), is
\be
\mathcal{I}=\frac{(8 a^4-50 r_0^2 a^2+15 r_0^4)a-15 r_0^2 [2 a^2+(2 r_{_+}-r_0)^2] (2 r_{_+}-r_0) \arctan\left(\frac{a}{r_{_+}}\right)}{2 a^2 r_{_+}^2 \left[(4a^2-3r_0^2)a+3 r_0^2 (2 r_{_+}-r_0) \arctan\left(\frac{a}{r_{_+}}\right)\right]}.
\ee
This expression is always negative and finite, as can be seen in Fig.~\ref{fig:kerr}. We conclude that the Kerr instanton is unstable for non-conformal perturbations. It is fortunate that this particular probe perturbation was able to identify the negative mode. Note that $\mathcal{I}(a=0)=-5/(7 r_0^2)\simeq -0.71 r_0^{-2}\gtrsim-0.76 r_0^{-2}$, the negative eigenmode of the Lichnerowicz operator found in \cite{Gross:1982cv}.


\subsection{\label{subsec:5Dmyersperry}Five dimensional Myers-Perry}

We will use the complexified form of the line element of the five dimensional Myers-Perry solution in the coordinates introduced in \cite{Chen:2006ea},
\be
\dd s^2 = (x+y) \left(\frac{\dd x^2}{4 X}+\frac{\dd y^2}{4Y}\right)-\frac{Y(\dd \tilde{\tau}-i x\dd\tilde{\phi}_1)^2}{y(x+y)}+\frac{(\dd \tilde{\tau}+i y\dd\tilde{\phi}_1)^2 X}{x (x+y)}-\frac{a_1^2 a_2^2}{x y} [\dd \tilde{\tau}-i x y \dd \tilde{\phi}_2-i(x-y)\dd \tilde{\phi}_1]^2,
\label{eq:myers_perry_5D_line}
\ee
where the coordinates $(\tilde{\tau},\tilde{\phi},\tilde{\psi})$ are related to the canonically defined coordinates $(\tau,\phi,\psi)$ via
\be
\begin{array}{cccc}
\tau=\tilde{\tau}+i(a_1^2+a_2^2)\tilde{\phi}_1+ia_1^2 a_2^2\tilde{\phi}_2, & \phi_1 = a_1 \tilde{\phi}_1+a_1 a_2^2 \tilde{\phi}_2 & \text{ and } & \phi_2 = a_2 \tilde{\phi}_1+a_2 a_1^2 \tilde{\phi}_2.
\end{array}
\ee
Also, $X$ and $Y$ are quadratic polynomials whose coefficients depend on the mass $r_0^2 =2 M$ and rotation parameters $(a_1,a_2)$,
\be
\begin{array}{ccc}
X = (x+a_1^2)(x+a_2^2)-r_0^2 & \text{ and } & Y = -(y-a_1^2)(y-a_2^2).
\end{array}
\ee
Again, to avoid a conical singularity at the horizon, located at the larger real root $x_{_+}$ of $X$, one must require the coordinate identification $(\tau,\phi_1,\phi_2)= (\tau+ \beta, \phi_1 - i \beta \Omega_1,\phi_2 - i \beta \Omega_2)$, where $\beta = 2\pi (x_{_+}+a_1^2)(x_{_+}+a_2^2)/[\sqrt{x_{_+}}X'(x_{_+})]$ and $\Omega_i = a_i/(x_{_+}+a_i^2)$ are the black hole inverse temperature and angular velocities, respectively. To compute the final integral we further need to specify the range of the coordinates on our manifold: $0\leq\tau<\beta$, $0\leq \phi_1,\;\phi_2 < 2\pi$, $x\geq x_{_+}$ and $\min(a_1^2,a_2^2)\leq y \leq \max(a_1^2,a_2^2)$.

\begin{figure}
\centering
\includegraphics[width = 8 cm]{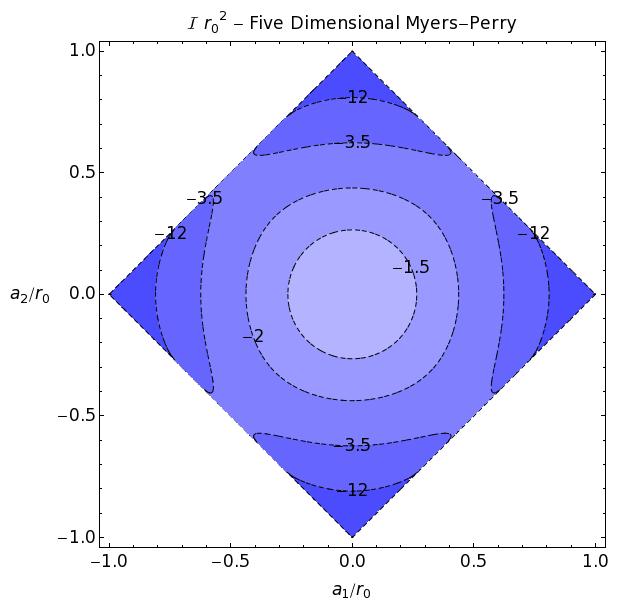}
\caption{\label{fig:myers_perry_5D} For the five dimensional Myers-Perry instanton, $\mathcal{I}$ is always negative. The dashed lines represent contours of constant $\mathcal{I}r_0^2$.}
\end{figure}

The final result for $\mathcal{I}$ is cumbersome and we will only graphically represent, in Fig.~(\ref{fig:myers_perry_5D}), the variation of $\mathcal{I}$ as a function of $a_1/r_0$ and $a_2/r_0$. The similarities with the Kerr black hole are evident. The shaded area is the parameter space where \eq{myers_perry_5D_line} represents a black hole, that is $a_1^2+a_2^2+2|a_1 a_2|\leq r_0^2$, where equality corresponds to extremality, and it coincides with the region where $\mathcal{I}$ is negative.
In contrast with the Kerr geometry, the five dimensional Myers-Perry geometry admits solutions with singular (infinite negative) $\mathcal{I}$, the corners in Fig.~(\ref{fig:myers_perry_5D}). These are the extremal solutions with rotation in a single plane, for which the horizon area is zero and there is a naked singularity. The leading behaviour near the singular points is well approximated by
\be
\mathcal{I}\simeq-\frac{274}{609(a_i\pm r_0)^2}.
\ee

In certain limits, $\mathcal{I}$ considerably simplifies, as for instance in the five dimensional Tangherlini solution ($a_1=a_2=0$) \cite{Tangherlini:1963bw}, in which case
\be
\mathcal{I}=-\frac{24}{19 r_0^2},
\ee
or in the singly-spinning Myers-Perry black hole limit ($a_2=0$),
\be
\mathcal{I}=-\frac{8 (160 a_1^{14}-1040 r_0^2 a_1^{12}+2794 r_0^4 a_1^{10}-3927 r_0^6 a_1^8+2925 r_0^8 a_1^6-870 r_0^{10} a_1^4-220 r_0^{12} a_1^2+315
   r_0^{14})}{21 r_0^4 (a_1^2-r_0^2)^2 (8 a_1^8-32 r_0^2 a_1^6+77 r_0^4 a_1^4-119 r_0^6 a_1^2+95 r_0^8)}.
\label{eq:myers_perry_5D_single}
\ee

We also point out the consistency of our results with the numerical treatment of the Gregory-Laflamme instability for rotating black string solutions \cite{Kleihaus:2007dg}. There is a correspondence between the Gregory-Laflamme instability of the black string and the existence of a thermodynamic negative mode of the black hole \cite{Reall:2001ag}. This correspondence implies that the negative eigenvalue is larger in magnitude if the threshold wavelength for the Gregory-Laflamme instability is smaller, and this is indeed what is found when comparing our results with \cite{Kleihaus:2007dg}.


\subsection{\label{subsec:emparanring}Singly-spinning black ring}
The complexified singly-spinning black ring line element is \cite{Astefanesei:2005ad}
\be
\dd s^2 = \frac{F(y)}{F(x)}\left[\dd \tau-i C R \frac{1+y}{F(y)}\dd \psi\right]^2+
\frac{R^2F(x)}{(x-y)^2}\left[-\frac{G(y)}{F(y)}\dd\psi^2-\frac{\dd y^2}{G(y)}+\frac{\dd x^2}{G(x)}+\frac{G(x)}{F(x)}\dd \phi^2\right],
\label{eq:ring1}
\ee
where
\be
\begin{array}{cccc}
F(\xi) = 1+\lambda \xi, & G(\xi) = (1-\xi^2)(1+\nu \xi), &
\text{ and } & C=\sqrt{\lambda(\lambda-\nu)\frac{1+\lambda}{1-\lambda}},
\end{array}
\ee
and the dimensionless parameters $\nu$ and $\lambda$ lie in the range $0<\nu\le\lambda<1$. As it stands, the line element defined above has conical singularities at $y=-1$, $x=-1$ and $x=1$. In order to remove the first two one must choose the periodicity of $\phi$ and $\psi$ to be
\be
\Delta \phi=\Delta\psi = 2\pi\frac{\sqrt{1-\lambda}}{1-\nu}.
\ee
The solution still has a conical singularity at $x=1$, and is often referred to in the literature as the unbalanced black ring. To obtain the physically acceptable black ring one must remove the conical singularity at $x=1$ by choosing $\lambda$ to be a function of $\nu$, leaving our solution dependent on two parameters $(R,\nu)$,
\be
\lambda = \frac{2\nu}{1+\nu^2}.
\label{eq:ring2}
\ee
The remaining parameters uniquely specify the mass and angular momentum of the black ring \cite{Emparan:2001wn}. The singly-spinning Myers-Perry solution is obtained by taking the limit $R\to0$ and $\lambda,\nu\to1$ in the unbalanced solution, while keeping fixed $a,\;r_0$ given by
\be
\begin{array}{ccc}
\displaystyle{r_0^2 = \frac{2 R^2}{1-\nu}} \quad & \text{ and }& \quad \displaystyle{a^2 = 2 R^2 \frac{\lambda-\nu}{(1-\nu)^2}.}
\end{array}
\label{eq:ring3}
\ee
Following \cite{Astefanesei:2005ad}, we note that the black ring instanton has another conical singularity located at the horizon $y=-1/\nu$, which is removed if we make the periodic identification $(\tau,\phi)=(\tau+\beta,\phi - i \beta \Omega)$, where $\beta = 4\pi R \sqrt{\lambda\nu(1+\lambda)}/[\sqrt{1-\lambda}(1+\nu)]$ and $\Omega = \sqrt{\lambda-\nu}/[R \sqrt{\lambda(1+\lambda)}]$. 

Again it is important to clearly state the range of coordinates in our manifold, which are given in the patch that we are interested in by $\tau\in[0,\beta)$, $\phi,\psi\in[0,2\pi\sqrt{1-\lambda}/(1-\nu))$, $y\in[-1/\nu,-1)$ and $x\in[-1,1)$.
\begin{figure}
\centering
\includegraphics[width = 8 cm]{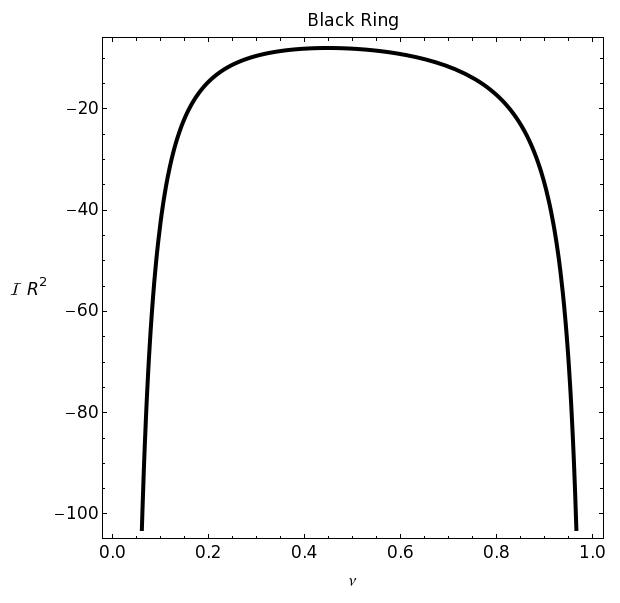}
\caption{\label{fig:ring}$\mathcal{I}$ is negative across the entire range.}
\end{figure}
We chose to determine $\mathcal{I}$ as a function of $(R,\nu,\lambda)$ and only latter imposing either relation (\ref{eq:ring2}) or the limit (\ref{eq:ring3}). The expressions for both $\phi_{ab}$ and $\mathcal{I}$ are complicated, but in the limit (\ref{eq:ring3}) $\mathcal{I}$ does reproduce \eq{myers_perry_5D_single}. Again we evaluated $\mathcal{I}$ using the relation (\ref{eq:ring2}), see Fig.~(\ref{fig:ring}), and we found that the black ring also exhibits an unstable behaviour against non-conformal perturbations.


\subsection{\label{subsec:6Dmyersperry}Six dimensional Myers-Perry}

In contrast to the previous cases, the moduli space of the six dimensional Myers-Perry solution is unbounded, in the sense that the black hole will exist for all values of one of the rotation parameters, if the other rotation parameter vanishes.

It is a challenging problem to compute $\mathcal{I}$ for this black hole, since we are dealing with a co-dimension three manifold, so that finding a suitable line element for the computation is important. Here we will complexify the line element presented in \cite{Chen:2006xh},
\begin{align}
\dd s^2 = & \frac{(r^2+y^2)(r^2+z^2)\dd r^2}{R}+\frac{(r^2+y^2)(y^2-z^2)\dd y^2}{Y}+\frac{(r^2+z^2)(z^2-y^2)\dd z^2}{Z} \nonumber \\
	  & + \frac{R}{(r^2+y^2)(r^2+z^2)}[\dd \tau -i (a_1^2-y^2)(a_1^2-z^2)\dd \tilde{\phi}_1-i (a_2^2-y^2)(a_2^2-z^2)\dd \tilde{\phi}_2]^2 \nonumber \\
	  & - \frac{Y}{(r^2+y^2)(y^2-z^2)}[\dd \tau -i (a_1^2+r^2)(a_1^2-z^2)\dd \tilde{\phi}_1-i (a_2^2+r^2)(a_2^2-z^2)\dd \tilde{\phi}_2]^2 \nonumber \\
	  & - \frac{Z}{(r^2+z^2)(z^2-y^2)}[\dd \tau -i (a_1^2+r^2)(a_1^2-y^2)\dd \tilde{\phi}_1-i (a_2^2+r^2)(a_2^2-y^2)\dd \tilde{\phi}_2]^2,
\label{eq:6Dmp1}
\end{align}
where $R$, $Y$ and $Z$ are quartic polynomials given by
\be
\begin{array}{cccc}
R(r) =(r^2+a_1^2)(r^2+a_2^2)-r_0^3 r,& Y(y) = -(a_1^2-y^2)(a_2^2-y^2) & \text{ and } & Z(z)=Y(z).
\end{array}
\ee
The coordinates $\tilde{\phi}_i$ are related with the canonically defined coordinates $(\phi_1,\phi_2)$ via
\be
\begin{array}{ccc}
\displaystyle{\tilde{\phi}_1=\frac{\phi_1}{a_1^2(a_1^2-a_2^2)}} \quad & \text{ and } & \quad \displaystyle{\tilde{\phi}_2=\frac{\phi_2}{a_2^2(a_2^2-a_1^2)}.}
\end{array}
\ee
The avoidance of conical singularities at the horizon, located at the largest root $r_{_+}$ of $R$, implies the coordinate identifications $(\tau,\phi_1,\phi_2)=(\tau+\beta,\phi_1 - i\beta\Omega_1,\phi_2 - i\beta\Omega_2)$, where $\beta=4 \pi r_0^3 r_{_+}/R'(r_{_+})$ and $\Omega_i = a_i/(r_{_+}^2+a_i^2)$. The region in moduli space in which the line element (\ref{eq:6Dmp1}) represents a black hole, dependent on the zeros of $R(r)$, is given by
\be
16 a_1^2 a_2^2 (a_1^2-a_2^2)^4-4 (a_1^6-33 a_2^2 a_1^4-33 a_2^4 a_1^2+a_2^6)r_0^6-27 r_0^{12}\leq0,
\ee
where the inequality is saturated for extremality and we have used the results compiled in \cite{312597}. This region of the moduli space obviously coincides with the region in which $\beta$ is positive semi-definite. The range of the variables $(\tau,r,y,z,\phi_1,\phi_2)$ is more involved: $\tau\in[0,\beta)$, $r\in[r_{_+},+\infty)$, $y\in[-b,b]$, $z\in[b,a]$ and $\phi_i\in [0,2\pi)$, where $b=\min(|a_1|,|a_2|)$ and $a=\max(|a_1|,|a_2|)$.

\begin{figure}
\centering
\includegraphics[width = 8 cm]{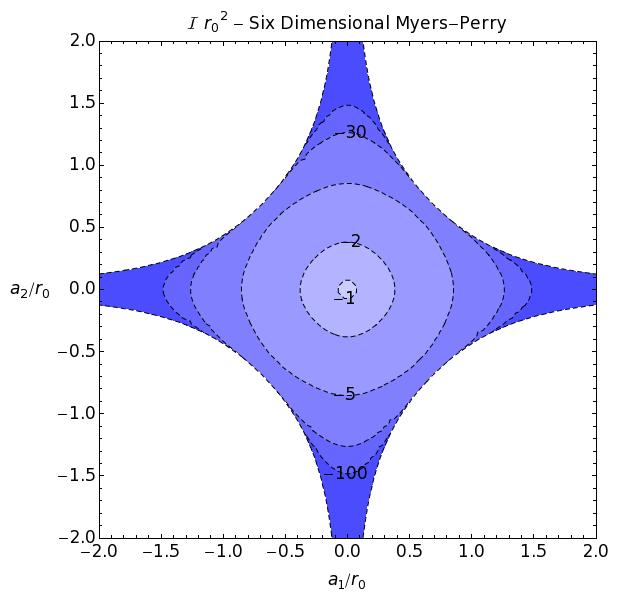}
\caption{\label{fig:myers_perry_6D}For the six dimensional Myers-Perry instanton, $\mathcal{I}$ is always negative. The dashed lines represent contours of constant $\mathcal{I}r_0^2$.}
\end{figure}

The numerical evaluation of $\mathcal{I}$ leads to a result very similar to the five dimensional case, see Fig.~(\ref{fig:myers_perry_6D}). $\mathcal{I}$ is negative for all values of $(a_1,a_2)$ that have a positive semi-definite $\beta$, \emph{i.e.} black holes.
The full expression considerably simplifies if we set both rotation parameters to zero,
\be
\mathcal{I}=-\frac{66}{169 r_0^2},
\ee
which corresponds to the six dimensional Tangherlini solution \cite{Tangherlini:1963bw}. Here, all extremal solutions have a non-divergent $\mathcal{I}$, except when one of the rotation parameters vanishes and the other becomes infinite. Near the singular points, $\mathcal{I}$ is well approximated by
\be
\mathcal{I}\simeq -\frac{747 a_i^{10}}{280 r_0^{12}}.
\ee


\subsection{\label{subsec:kerrAdS}Kerr-AdS black hole}

The Kerr-AdS instanton is given by
\be
\dd s^2 = \frac{\Delta}{\Sigma^2} \left( \dd\tau -i \frac{a}{\Xi} \sin^2\theta \dd\phi \right)^2 + \Sigma^2 \left( \frac{\dd r^2}{\Delta} +\frac{\dd\theta^2}{\Delta_\theta} \right)-\frac{\Delta_\theta \sin^2\theta}{\Sigma^2} \left(a \dd\tau-i \frac{r^2+a^2}{\Xi} \dd\phi \right)^2
\label{eq:kerrAdS1}
\ee
with
\be
\begin{array}{ccccc}
\Delta=(r^2+a^2)(1+r^2\ell^{-2}) -r_0 r, & \Sigma^2=r^2 +a^2 \cos^2\theta, & \Delta_\theta=1-a^2\ell^{-2} \cos^2{\theta}, \text{ and } & \Xi=1-a^2\ell^{-2},
\end{array}
\ee
where $\ell$ is the curvature radius of AdS and is related to the cosmological constant as $\ell^2 =3/|\Lambda|$. The Kerr solution is recovered in the usual limit $\ell\to+\infty$. The line element (\ref{eq:kerrAdS1}) only represents a physical solution for $|a|<\ell$, being singular in the limit $|a| \to \ell$ for which the 3-dimensional Einstein universe at infinity rotates at the speed of light \cite{Hawking:1998kw}.

The avoidance of a conical singularity at the horizon, located at $r= r_{_+}$, the largest root of $\Delta$, requires, as in the Kerr case, the coordinate identification $(\tau,\phi)= (\tau+ \beta, \phi - i \beta \Omega_\mathrm{rot})$. Here,
\be
\beta=\frac{4\pi(r_{_+}^2 +a^2)}{r_{_+}(1+a^2\ell^{-2} + 3 r_{_+}^2 \ell^{-2} - a^2r_{_+}^{-2})}
\ee
and
\be
\Omega_\mathrm{rot} = \frac{a(1- a^2\ell^{-2})}{r_{_+}^2 + a^2}.
\ee
The angular velocity $\Omega_\mathrm{rot}$ here is measured relative to a frame rotating at infinity. It is convenient to choose a coordinate system that is not rotating at infinity ($ \tilde{\phi} \to \phi - i a \ell^{-2} \tau$), for which the angular coordinate identification depends instead on the angular velocity
\be
\Omega = \frac{a(1+ r_{_+}^2\ell^{-2})}{r_{_+}^2 + a^2}.
\ee
The reason for this is that $\Omega_\mathrm{rot}$ is not the appropriate thermodynamic variable. It cannot be used to formulate the first law of thermodynamics \cite{Gibbons:2004ai}. The issue is irrelevant in this Section, as we are interested only in the coordinate independent quantity $\mathcal{I}$, but will be very important for the thermodynamic interpretation of the results.

The expression for the TT perturbation $\phi_{ab}$ is remarkably simple, and is given by
\be
\phi^a_{\phantom{a}b}=\frac{r_0^2}{2 \Sigma^4}\left[
\begin{array}{cccc}
\displaystyle{1+\frac{2 a^2 \sin ^2\theta}{\Sigma^2}} & 0 & 0 & \displaystyle{-\frac{2 i a(r^2+a^2) \sin ^2\theta}{\Sigma^2\Xi}} \\
 0 & 1 & 0 & 0 \\
 0 & 0 & -1 & 0 \\
\displaystyle{-\frac{2 i a\Xi}{\Sigma^2}} & 0 & 0 & \displaystyle{-\Big(1+\frac{2 a^2 \sin ^2\theta}{\Sigma^2}\Big)}
\end{array}
\right].
\ee
This perturbation is clearly traceless, and in the limit $\ell^{-1}\to0$ reduces to \eq{kerrprobe}. In the absence of rotation we know, through \cite{Prestidge:1999uq}, that the negative mode ceases to exist exactly when the specific heat becomes positive. This transition occurs at $\ell_c = (3\sqrt{3}/4) r_0$, the specific heat being positive for $\ell<\ell_c$. This is similar to what happens when one considers Schwarzschild in a finite cavity \cite{York:1986it}.  We expect that, in the absence of rotation, $\mathcal{I}$ will change sign for $\ell_\mathcal{I}>\ell_c$, because $\phi_{ab}$ is not the negative eigenmode. In fact, $\mathcal{I}$ changes sign at $\ell_\mathcal{I}\simeq1.655 r_0$, a value of $\ell$ bigger, but close to $\ell_c$. Classically, we also know that AdS black holes are unstable to superradiant modes whenever $\Omega \ell>1$ \cite{Hawking:1999dp}.
\begin{figure}
\centering
\includegraphics[width = 8 cm]{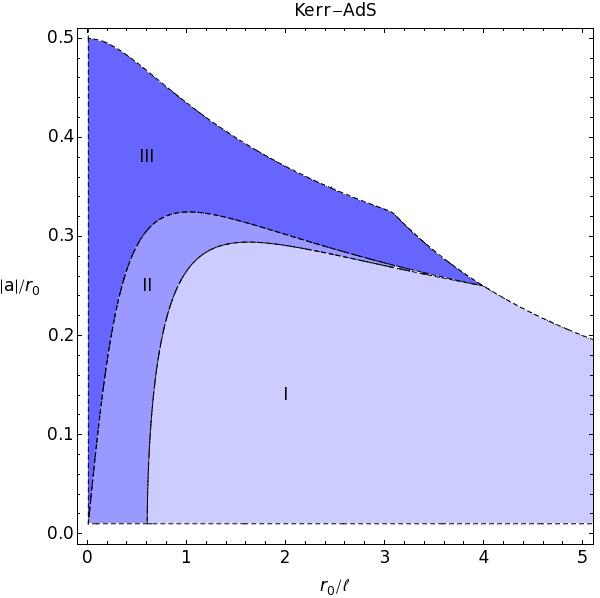}
\caption{\label{fig:kerr_ads}Phase diagram of the Kerr-AdS black hole.}
\end{figure}

In Fig.~(\ref{fig:kerr_ads}), we represent a phase diagram for the Kerr-AdS black hole. In region I, $\mathcal{I}$ is positive, so we conclude nothing apart from the consistency with known results in the non-rotating limit \cite{Prestidge:1999uq}. The region of most interest is region II. Here the black holes are not superradiant, and may be classically stable, but $\mathcal{I}$ is still negative, meaning that black holes in this region are thermodynamically unstable. Region III corresponds to superradiant and thermodynamically unstable black holes. Finally, the white region corresponds to unphysical instantons, beyond extremality or the bound $|a|<\ell$. It is a curious fact that the line that divides non-superradiant black holes from superradiant black holes meets the contour line corresponding to $\mathcal{I}=0$ at  $|a|=\ell$.

\section{\label{sec:interpretation}Thermodynamic interpretation}


\subsection{\label{subsec:partstab}Ensembles and stability}
The partition functions studied here have boundary conditions that impose periodicities both in time and in the rotation angles. This corresponds to fixing the temperature $T= \beta^{-1}$ and the angular velocities $\Omega_i$, \emph{i.e.} to the grand-canonical ensemble. The canonical ensemble would require fixing the angular momenta $J^i$, which includes specifying normal derivatives of the metric on the boundary. We expect a correspondence between the stability of the ensemble and the problem of non-conformal perturbations about the gravitational instanton. If there is a negative mode in the path integral, the partition function is ill-defined and the ensemble should not be stable.

In this section, we recall the conditions for the thermodynamic stability of a system described by the grand-canonical ensemble. The laws of thermodynamics imply that, for a system in equilibrium at given temperature and angular velocities, any deviation from equilibrium, not necessarily small, obeys \cite{Landau}
\be
\delta E - T \delta S - \Omega_i \delta J^i > 0.
\ee
Here, the angular velocities $\Omega_i$ play the usual role of chemical potentials and the angular momenta $J^i$ are the respective charges (conserved in the canonical ensemble). Expanding $\delta E(S,J^i)$ to second order and using the first law of thermodynamics,
\be
\label{1stlaw}
\dd E=T\dd S +\Omega_i \dd J^i,
\ee
we arrive at the conclusion that the stability relies only on the Hessian of the energy
\be
{g^{(W)}}_{\mu \nu} = {\dderf{E}{x^\mu}{x^\nu}}, \qquad x^\mu=(S,J^i),
\ee
being positive definite. This Hessian matrix defines the so-called Weinhold metric \cite{Weinhold}. Its inverse is given by
\be
{g^{(W)}}^{\mu \nu} = -{\dderf{G}{y_\mu}{y_\nu}}, \qquad y_\mu=(T,\Omega_i),
\ee
where $G$ is the Gibbs free energy,
\be
G= E - T S - \Omega_i J^i.
\ee
The positivity of the inverse Weinhold metric gives the same stability condition. Notice that the first law is equivalent to
\be
\dd G= -S\dd T - J^i \dd\Omega_i,
\ee
so that the coordinate systems $x^\mu$ and $y_\mu$ are related by $y_\mu = \partial E/\partial x^\mu$ and $x^\mu = -\partial G/\partial y_\mu$.
A third alternative is the positivity of the Ruppeiner metric \cite{Ruppeiner},
\be
{g^{(R)}}_{\mu \nu}= -{\dderf{S}{u^\mu}{u^\nu}}, \qquad u^\mu=(E,J^i),
\ee
which is easily shown to be conformal to the Weinhold metric,
\be
\dd s_R^2 = -{\dderf{S}{u^\mu}{u^\nu}} \dd u^\mu \dd u^\nu = \beta {\dderf{E}{x^\mu}{x^\nu}} \dd x^\mu \dd x^\nu = \beta \dd s_W^2.
\ee

Let us now look at how the stability conditions relate to the usual linear response functions, like the specific heat. First, notice that
\be
-{\dderf{G}{y_\mu}{y_\nu}} = \left[
\begin{array}{cc}
\beta C_\Omega & \eta^j \\
\eta^i & \epsilon^{ij}
\end{array}
\right],
\ee
where
\be
C_\Omega = T \left( {\der{S}{T}} \right)_\Omega
\ee
is the specific heat at constant angular velocities (all $\Omega_i$ fixed). The isothermal differential moment of inertia tensor is
\be
\epsilon^{ij} = \left( {\der{J^i}{\Omega_j}} \right)_T = \epsilon^{ji}.
\ee
There is also the vector
\be
\eta^{i} = \left( {\der{S}{\Omega_i}} \right)_T = \left( {\der{J^i}{T}} \right)_\Omega,
\ee
where the second equality, given by the symmetry of the Hessian matrix, corresponds to a Maxwell relation. The $\eta^{i}$ dependence can be dealt with if we use the relation between the specific heat at constant angular velocities $C_\Omega$ and the specific heat at constant angular momenta $C_J$, defined as
\be
C_J = T \left( {\der{S}{T}} \right)_J,
\ee
which gives
\be
\label{cocj}
C_\Omega = C_J + T (\epsilon^{-1})_{ij} \eta^{i} \eta^{j}.
\ee
Hence
\be
\label{eq:ds2diag}
\dd s_W^2 = -{\dderf{G}{y_\mu}{y_\nu}} \dd y_\mu \dd y_\nu = \beta C_J \dd y_0^2 + \epsilon^{ij} \omega_i \omega_j,
\ee
where $\omega_i=\dd y_i + (\epsilon^{-1})_{ij} \eta^{j} \dd y_0$.

From \eq{ds2diag}, we conclude that the thermodynamic stability is given by $C_J$ and the spectrum of $\epsilon^{ij}$, the condition being
\be
C_J >0 \qquad \mathrm{and} \qquad \mathrm{spec} (\epsilon^{ij})>0.
\ee


\subsection{\label{subsec:stabnegmodes}Stability and negative modes}
We expect that the partition function correctly reproduces the thermodynamic features of a system. In particular, we expect that the partition function, which we associate with a system in equilibrium, has some sort of pathology if thermodynamic stability does not hold. In this section, we will extend a result by Reall \cite{Reall:2001ag} (based on previous contributions \cite{PhysRevLett.61.1336,Prestidge:1999uq}) on the relation between thermodynamic stability and the existence of a negative mode. He showed that a negative specific heat implied the existence of a negative mode in the action, in the case of the canonical ensemble. We wish to extend this argument to the grand-canonical ensemble. The converse result - to show that negative modes can only originate from thermodynamic instability - has not been proven, and we have found a counter-example (Section~\ref{subsubsec:kerr_ads}).

The proof has two steps. The first step is to consider a path of geometries, intersecting a black hole saddle-point solution, on which the action functional has a certain dependence. That dependence implies an explicit relation between negative modes and thermodynamic stability. The second step is to show that the construction is possible, \emph{i.e.} that such a path indeed exists.

Consider paths of geometries, parametrised by variables $z^\mu$ for given temperature $T$ and angular velocities $\Omega_j$, for which the action takes the form
\be
I(z^\mu,T,\Omega_j) = \beta \tilde{E}(z^\mu) - \tilde{S}(z^\mu) - \beta \Omega_i \tilde{J}^i(z^\mu).
\ee
These paths intersect black hole solutions with temperature $T$ and angular velocities $\Omega_j$, which are given by $z^\mu=z^\mu_*(T,\Omega_i)$. For the black hole geometries, $I=\beta G$, so that
\be
\tilde{E}(z^\mu_*) = E(T,\Omega_i), \qquad \tilde{S}(z^\mu_*) = S(T,\Omega_i), \qquad \tilde{J}^i(z^\mu_*) = J^i(T,\Omega_j).
\ee
The index in $z^\mu$ has the same range as the index in $y_\mu=(T,\Omega_i)$, the coordinates used in the last section to define the inverse Weinhold metric. The invertibility of $z^\mu_*(y_\nu)$ is assumed in the following. The argument works because the functions that only depend on $z^\mu$ can be thought of as depending on $y_\mu(z^\nu)$.

At saddle points of $I$, \emph{i.e.} for black hole solutions,
\be
\label{Iconst}
\der{I}{z^\mu}(z^\alpha_*) = \der{y_\nu}{z^\mu} \left( \beta \der{E}{y_\nu}  - \der{S}{y_\nu} - \beta \Omega_i \der{J^i}{y_\nu}\right) =0,
\ee
which implies the first law of thermodynamics (\ref{1stlaw}). The Hessian of $I$ at these points takes the form
\be
{\dderf{I}{z^\mu}{z^\nu}} (z^\alpha_*) = \beta
\left[
\begin{array}{cc}
\displaystyle{\der{T}{z^\mu}} & \displaystyle{\der{\Omega_i}{z^\mu}}
\end{array}
\right]
\left[
\begin{array}{cc}
\beta C_\Omega & \eta^j \vspace{2 mm} \\
\eta^i & \epsilon^{ij}
\end{array}
\right]
\left[
\begin{array}{c}
\displaystyle{\der{T}{z^\nu}}  \vspace{2 mm} \\  \displaystyle{\der{\Omega_i}{z^\nu}}
\end{array}
\right].
\ee
This result means that the negative modes of $I$ given by this class of geometries have their origin in the non-positivity of the Weinhold metric,
\be
\delta I = \frac{1}{2} {\dderf{I}{z^\mu}{z^\nu}} \Big|_{z_*} \hspace{-1.5 mm} \delta z^\mu \delta z^\nu = \frac{\beta}{2} \, {\delta s_W}^2.
\ee
Notice that the path integral includes geometries that might not fall into this class. That is the reason why the proof of the converse result does not follow from the construction above.

It would now be necessary to show that a path of geometries satisfying (\ref{Iconst}) is possible. Fortunately, this problem was addressed already by Brown, Martinez and York \cite{Brown:1990di} for the Kerr black hole, and the extension to higher dimensional cases should present no difficulty. They consider four dimensional axisymmetric geometries which satisfy the Einstein constraints and have the appropriate boundary conditions on the wall of a finite cavity. In a such a cavity, it is possible to have thermodynamically stable black holes which are asymptotically flat (in this work, we consider the technically simpler boundary conditions imposed at infinity). The final result of the construction, their expression (43) for the gravitational action, exactly reproduces our assumption (\ref{Iconst}).


\subsection{\label{subsec:discussion}Particular cases}
We have seen in Section~\ref{subsec:partstab} that the linear response functions that characterise the stability are the specific heat at constant angular momenta and the moment of inertia tensor. In this section, we will calculate such quantities and interpret the results obtained with $\mathcal{I}$.


\subsubsection{\label{subsubsec:kerr}Kerr black hole}
\begin{figure}
\centering
\includegraphics[width = 8 cm]{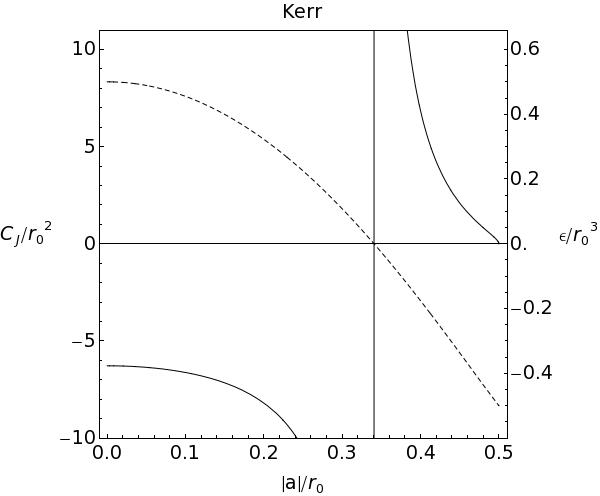}
\caption{\label{fig:kerr_perme}Graphic representation of $C_J$ (solid line with axis on the left) and $\epsilon$ (dashed line with axis on the right).}
\end{figure}

Here the specific heat is given by
\be
C_J = \frac{2 \pi(r_{_+}^2-a^2)(r_{_+}^2+a^2)^2}{3 a^4+6 r_{_+}^2 a^2-r_{_+}^4},
\label{eq:kerr_heat}
\ee
where $C_J$ is negative for small values of $|a|$, but becomes positive for $|a|>\sqrt{2\sqrt{3}-3}r_0/2$. The moment of inertia can be expressed as
\be
C_J\epsilon=-\frac{\pi (r_{_+}^2+a^2)^2(r_{_+}^2-a^2)}{r_{_+}},
\label{eq:kerr_perme}
\ee
which is negative definite for black hole solutions. We conclude that, when $C_J$ is negative, $\epsilon$ is positive, and vice versa. From the grand-canonical point of view, the Kerr black hole is thermodynamically unstable, justifying why $\mathcal{I}$ was negative for all values of $|a|<r_0/2$. We have plotted both quantities in Fig.~(\ref{fig:kerr_perme}). Furthermore, in the absence of rotation, $\epsilon$ is positive, meaning that the Schwarzschild black hole is stable against perturbations in the angular velocity.


\subsubsection{\label{subsubsec:myers_perry_5D}Five dimensional Myers-Perry}
\begin{figure}
\centering
\includegraphics[width = 8 cm]{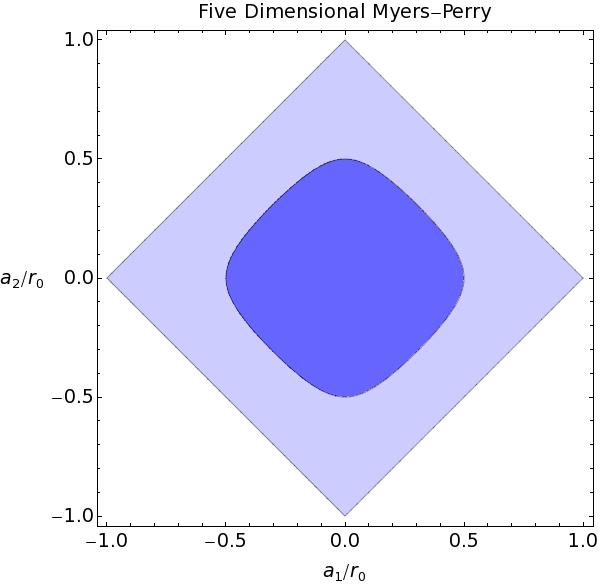}
\caption{\label{fig:myers_perry_5D_perme}The darker region indicates the region in the moduli space for which $C_J$ is negative, and the lighter region indicates where $\epsilon$ is not positive definite.}
\end{figure}

The specific heat at constant angular momenta and the moment of inertia tensor associated with the five dimensional Myers-Perry instanton are respectively given by
\be
C_J=\frac{2 \pi(x_{_+}+a_1^2)^2(x_{_+}+a_2^2)^2(x_{_+}^2-a_1^2 a_2^2)}{a_1^2 a_2^2 [5 a_1^2 a_2^2+9(a_1^2+a_2^2) x_{_+}+20 x_{_+}^2] x_{_+}^{1/2}+3(a_1^2+a_2^2) x_{_+}^{7/2}-x_{_+}^{9/2}},
\label{eq:myers_perry_5D_heat}
\ee
and
\be
\epsilon=
\left[
\begin{array}{cc}
-\frac{(x_{_+}+a_1^2)(x_{_+}+a_2^2)[3 a_2^2 a_1^4+3 x_{_+}(a_2^2-x_{_+}) a_1^2+x_{_+}^3]}{2 x_{_+} a_1^2 a_2^2-2
   x_{_+}^3} & -\frac{a_1 a_2 (x_{_+}+a_1^2)(x_{_+}+a_2^2)(a_1^2 a_2^2-3 x_{_+}^2)}{x_{_+} a_1^2 a_2^2-x_{_+}^3}
\\
\\
-\frac{a_1 a_2 (x_{_+}+a_1^2)(x_{_+}+a_2^2)(a_1^2 a_2^2-3 x_{_+}^2)}{x_{_+} a_1^2 a_2^2-x_{_+}^3} &
   -\frac{(x_{_+}+a_1^2)(x_{_+}+a_2^2)[x_{_+}^3-3 a_2^2 x_{_+}^2+3 a_1^2 a_2^2(a_2^2+x_{_+})]}{2 x_{_+} a_1^2
   a_2^2-2 x_{_+}^3}
\end{array}
\right].
\label{eq:myers_perry_5D_perme}
\ee
We have plotted the region in parameter space for which $\epsilon$ is not positive definite, and the region for which the specific heat at constant angular momenta is negative, see Fig. (\ref{fig:myers_perry_5D_perme}). As in the Kerr case, the two regions are complementary, meaning that the specific heat becomes positive when the moment of inertia ceases to be positive definite. This indicates that the Myers-Perry black hole is thermodynamically unstable in the grand-canonical ensemble, in agreement with the result obtained for $\mathcal{I}$. 


\subsubsection{\label{subsubsec:black_ring_5D}Singly-spinning black ring}
\begin{figure}
\centering
\includegraphics[width = 8 cm]{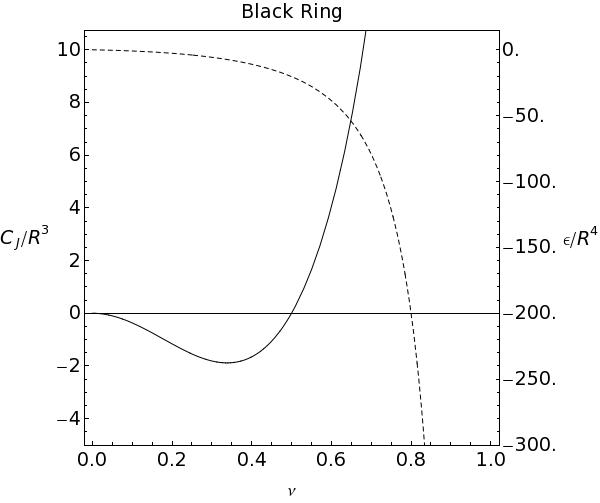}
\caption{\label{fig:ring_perme}Graphic representation of $C_J$ (solid line with axis on the left) and $\epsilon$ (dashed line with axis on the right).}
\end{figure}

The specific heat of the singly-spinning black ring is given by
\be
C_J=-\frac{6\sqrt{2}\pi^2 R^3(1-2\nu)\nu^2}{(1-\nu)(1+\nu^2)^{3/2}(2+\nu^2)},
\label{eq:black_ring_heat}
\ee
where $C_J$ is negative for small values of $\nu$, but becomes positive for $\nu>1/2$. However, the moment of inertia, given by
\be
\epsilon = -\frac{\pi R^4(1+\nu)^2(2+\nu ^2)\nu}{(1-\nu)^2(1+\nu ^2)^2},
\label{eq:black_ring_perme}
\ee
is negative for all values of $\nu$. We have plotted both linear response functions in Fig.~(\ref{fig:ring_perme}). The fact that $\epsilon$ is negative definite renders the black ring unstable in the grand-canonical ensemble, confirming the results obtained with $\mathcal{I}$. There is however a difference relative to all other black holes that we have analysed. In the ring case, the regions where $C_J<0$ and $\epsilon<0$ are not complementary. As a result, the specific heat at constant angular velocity,
\be
C_{\Omega}=-\frac{\sqrt{2} \pi ^2 R^3 \nu ^2 (\nu  (\nu +2)-2)}{(\nu -1) \left(\nu ^2+1\right)^{3/2}},
\ee
does not have a definite sign, becoming positive for $\nu>\sqrt{3}-1$. This means that $C_\Omega$ alone does not always identify the instability of the grand-canonical ensemble. We also point out that, in the extremal limit $\nu=1$, both the specific heat and the moment of inertia are divergent. This divergence is most likely related with the divergence of the energy and angular momentum at extremality.


\subsubsection{\label{subsubsec:myers_perry_6D}Six dimensional Myers-Perry}
The specific heat at constant angular momenta is given by
\be
C_{J}=-\frac{2 \pi(r_{_+}^2+a_1^2)^2(r_{_+}^2+a_2^2)^2[a_1^2 a_2^2-(a_1^2+a_2^2) r_{_+}^2-3 r_{_+}^4]}{5
   a_1^4 a_2^4+14 a_1^2 a_2^2(a_1^2+a_2^2) r_{_+}^2-(3 a_1^4-44 a_2^2 a_1^2+3 a_2^4) r_{_+}^4+6 (a_1^2+a_2^2) r_{_+}^6-3r_{_+}^8},
\label{eq:myers_perry_6D_heat}
\ee
whereas the moment of inertia tensor is $\epsilon = (r_{_+}^2+a_1^2)(r_{_+}^2+a_2^2)\tilde{\epsilon}$, where
\be
\tilde{\epsilon} =\left[
\begin{array}{cc}
\frac{3 r_{_+}^6-(a_2^2+6 a_1^2) r_{_+}^4+3(a_1^4+2 a_2^2 a_1^2) r_{_+}^2+3 a_1^4 a_2^2}{2 r_{_+}[3r_{_+}^4-(a_1^2+a_2^2) r_{_+}^2-a_1^2 a_2^2]} & -\frac{a_1 a_2 [7 r_{_+}^4-(a_1^2+a_2^2) r_{_+}^2-a_1^2a_2^2]}{r_{_+}[3 r_{_+}^4-(a_1^2+a_2^2) r_{_+}^2-a_1^2 a_2^2]}
\\
\\
 -\frac{a_1 a_2[7 r_{_+}^4-(a_1^2+a_2^2) r_{_+}^2-a_1^2 a_2^2]}{r_{_+}[3 r_{_+}^4-(a_1^2+a_2^2) r_{_+}^2-a_1^2a_2^2]} & \frac{3 r_{_+}^6-(a_1^2+6 a_2^2) r_{_+}^4+3 (a_2^4+2 a_1^2 a_2^2) r_{_+}^2+3 a_1^2 a_2^4}{2 r_{_+}[3r_{_+}^4-(a_1^2+a_2^2) r_{_+}^2-a_1^2 a_2^2]}
\end{array}
\right].
\ee

\begin{figure}
\centering
\includegraphics[width = 8 cm]{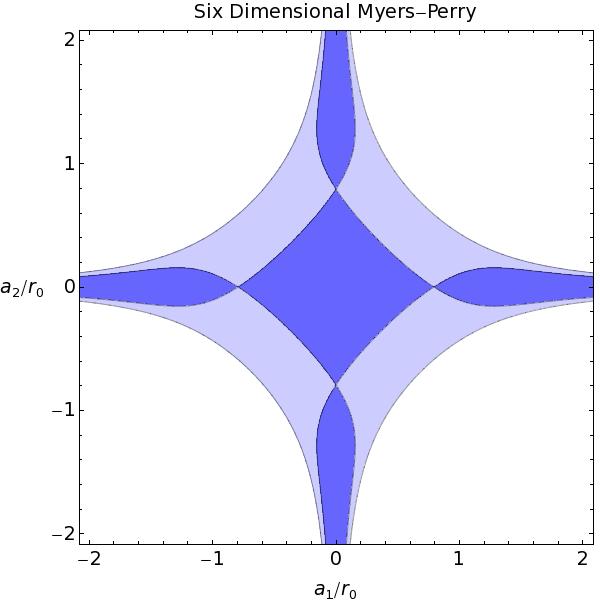}
\caption{\label{fig:myers_perry_6D_perme}The darker region indicates the region in the moduli space for which $C_J$ is negative. The region in which $\epsilon$ is not positive definite includes the lighter region.}
\end{figure}

We have again plotted $C_J$ as a function of the normalised rotation parameters $a_i/r_0$, see Fig.~(\ref{fig:myers_perry_6D_perme}). The similarities with the five dimensional Myers-Perry solution are evident. The region in which $C_J$ is negative is complementary to the region where one of the eigenvalues of $\epsilon$ is negative. We are again led to conclude that the six dimensional Myers-Perry solution is thermodynamically unstable in the grand-canonical ensemble, in agreement with the results obtained for $\mathcal{I}$.


\subsubsection{\label{subsubsec:kerr_ads}Kerr-AdS black hole}
\begin{figure}
\centering
\includegraphics[width = 8 cm]{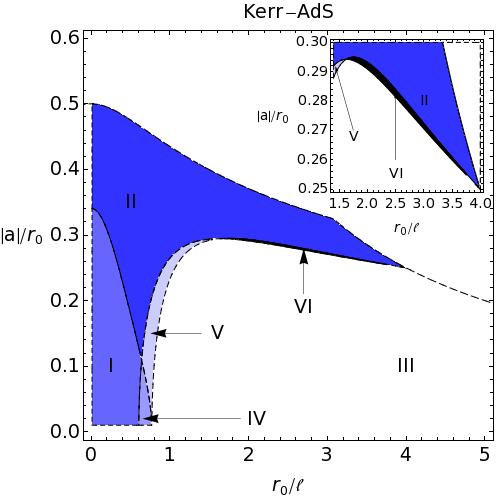}
\caption{\label{fig:kerr_ads_perme}Phase diagram of the Kerr-AdS black hole. In region III, the linear response functions $C_J$ and $\epsilon$ are positive and we find $\mathcal{I}>0$. In regions II and V, $\epsilon<0$ and $C_J>0$, while, in regions I and IV, $C_J<0$ and $\epsilon>0$. We find a negative mode, \emph{i.e.} $\mathcal{I}<0$, in regions I, II and VI. The thin region VI, amplified in the top-right corner, represents a puzzling feature since we find a negative mode when thermodynamic stability was expected.}
\end{figure}

The Kerr-AdS case is richer but also more subtle. We can calculate the specific heat and moment of inertia using the expressions in \cite{Gibbons:2004ai}, for which we obtain
\be
C_J=-\frac{2 \pi  \ell^4 (a^2+r_+^2)^2[(r_+^2-\ell ^2) a^2+r_+^2 (\ell ^2+3 r_+^2)]}{(a^2-\ell^2) [(3 a^4+6 r_+^2 a^2-r_+^4) \ell^4+(a^6+13 r_+^2 a^4+23 r_+^4 a^2+3 r_+^6) \ell^2+a^2 r_+^2(a^2+3 r_+^2)^2]}
\ee
and
\be
C_J\;\epsilon =-\frac{\pi\ell^8(a^2+r_+^2)^3 [(r_+^2-\ell^2) a^2+r_+^2(3 r_+^2+\ell^2)]}{r_+ (a^2-\ell^2)^4[(r_+^2+\ell^2) a^2-r_+^2(3 r_+^2-\ell^2)]}.
\ee
The parameter region for which the last expression is negative is composed, as in the previous Kerr and Myers-Perry cases, of two complementary regions. The specific heat at constant angular momentum is negative in one of them - regions I and IV in Fig.~(\ref{fig:kerr_ads_perme}) - and the moment of inertia is negative in the other - regions II and V in Fig.~(\ref{fig:kerr_ads_perme}). However, there is a region where both are positive - regions III and VI in Fig.~(\ref{fig:kerr_ads_perme}) - which contains, in the non rotation limit $a=0$, the known stable phase of large Schwarzschild-AdS black holes.

The areas IV, V and VI in Fig.~(\ref{fig:kerr_ads_perme}) identify the gap between a negative mode, found in Section~\ref{subsec:kerrAdS}, and the line of critical stability, found using the thermodynamics formulae. On the left, we have a strip in which we expect thermodynamic instability, based on the linear response functions $C_J$ (in region IV) and $\epsilon$ (in region V), but find no negative mode. This is not surprising because our perturbation is not a negative eigenmode and we do not expect it to identify the entire region of instability, which in the non-rotation limit $a=0$ persists until the specific heat becomes positive. There is however a very thin strip, region VI, amplified on the top-right corner of Fig.~(\ref{fig:kerr_ads_perme}), for which we detect a negative mode when no instability was expected in terms of the thermodynamic response functions. We find $\mathcal{I}<0$ but the Weinhold metric indicates stability. This is a puzzling feature as it seems to contradict our expectations that the partition function exactly reproduces the known thermodynamics.

The statement of the puzzle above immediately indicates a possible explanation for this small discrepancy. It is well known that the partition function, canonical or grand-canonical, which accounts for statistical physics, does not always reproduce thermodynamics. The `thermodynamic limit' arises only as a macroscopic approximate description. In the grand-canonical ensemble, the partition function reads
\be
Z(T,\Omega) = \int \dd E \dd J \, \omega(E,J) \, e^{-\beta(E-\Omega J)},
\ee
where $\omega(E,J)$ is the density of states with given energy $E$ and angular momentum $J$. The variance of the statistical distribution is given by
\be
\label{E2}
\frac{ \langle E^2 \rangle - \langle E \rangle^2 }{ \langle E \rangle^2 } = \frac{T^2}{\langle E \rangle^2} \left( \der{\langle E \rangle}{T} \right)_{(\beta \Omega)},
\ee
\be
\label{J2}
\frac{ \langle J^2 \rangle - \langle J \rangle^2 }{ \langle J \rangle^2 } = \frac{T}{\langle J \rangle^2} \left( \der{\langle J \rangle}{\Omega} \right)_{T}.
\ee
These variances are not small in most of the parameter space and diverge along the critical line of stability if we take $\langle E \rangle$ and $\langle J \rangle$ to be the thermodynamic quantities $E$ and $J$ used before. The derivative factor on the RHS of (\ref{J2}) is then the moment of inertia $\epsilon$, and the quantity held fixed in the derivative on the RHS of (\ref{E2}) is related to the so-called fugacity in the statistical mechanics of systems of particles. We saw that the RHS of (\ref{J2}) changes sign across the divergence on the line of critical stability, but the RHS of (\ref{E2}) is always positive in spite of the divergence.

In the region close to a phase transition, a careful study of the contribution of fluctuations is essential. If the thermodynamic description breaks down near a critical point, the precise determination of quantities like the critical temperature $T_c$ requires the contribution of higher order corrections to the partition function. A well understood example is the two-dimensional zero-field Ising model, which can be solved exactly. The exact critical temperature corresponds to almost half of the na\"ive leading order result, and the first correction is already very important (see Fig.~(9.5) of \cite{Stanley}; the van der Waals gas is another good example described in this reference). In our case, the one loop contribution is the first quantum correction and thus it is not surprising that the critical line, as predicted by the zero-th order quantities, will suffer a small correction.

We must point out that a linear response function, namely the specific heat at constant charge, also diverges in other known cases and the negative mode still disappears when expected. This happens for the Schwarzschild-AdS black hole, for which the specific heat diverges at $\ell_c = (3\sqrt{3}/4) r_0$, where the negative mode has numerically been shown to disappear \cite{Prestidge:1999uq}. It also occurs in the case of the Reissner-Nordstr\"om black hole, for which the specific heat at constant electromagnetic charge $C_Q$ diverges at $Q=\sqrt{3}M/2$, where the negative mode has analytically been shown to disappear \cite{monteiro:064006} (the latter work focuses on the canonical ensemble, so that both the stability and the validity of the thermodynamic limit are given by $C_Q$). In the Kerr-AdS case studied here, we are looking at the grand-canonical ensemble. The specific heat at constant angular momentum $C_J$ is finite at the transition, which is signalled instead by the quantities on the RHS of (\ref{E2}) and (\ref{J2}). Since
\be
\left( \der{\langle E \rangle}{T} \right)_{(\beta \Omega)} = \left( \der{\langle E \rangle}{T} \right)_{\langle J \rangle} + \beta \left( \der{\langle J \rangle}{\Omega} \right)_{T}  \left( \der{\langle E \rangle}{\langle J \rangle} \right)_{T}^2,
\ee
or, taking $\langle E \rangle$ and $\langle J \rangle$ to be the thermodynamic quantities $E$ and $J$ used before,
\be
\left( \der{E}{T} \right)_{(\beta \Omega)} = C_J + \beta \,\epsilon  \left(\der{E}{J}\right)_{T}^2.
\ee
Now, $C_J$ is divergent and $\epsilon=0$ along the line separating the regions I-IV and II-V in Fig.~(\ref{fig:kerr_ads_perme}). This corresponds to the transition in the canonical ensemble. In the grand-canonical ensemble, this line has no special interest and the LHS of the latter expression is still finite, because the divergence in $(\partial{E}/\partial{J})_{T}$ cancels the divergence in $C_J$. On the other hand, the line of critical stability, separating regions III-VI from the rest, is marked by a divergence in $\epsilon$ that also causes the divergence on the LHS, while the other quantities are finite. The exception is the single point for which $a=0$, the Schwarzschild-AdS case. There, the transition is signalled by the divergence of the specific heat since the moment of inertia is actually finite for $a=0$, $\epsilon = r_+^3/2$.

We conclude that the phase transition in the grand-canonical ensemble can be traced back to the divergence of the moment of inertia $\epsilon$ only. We speculate that this transition may have different quantum properties such that the critical stability line is corrected by the one loop contribution, while the same does not happen in the canonical case with the specific heat at constant charge, in the cases known.

To complete the discussion of the transitions, let us consider the critical exponents. Notice that these too can be corrected by higher order contributions to the partition function, but we are only able to study them here for the zero-th order semiclassical result corresponding to our thermodynamics formulae.

\begin{figure}
\centering
\includegraphics[width = 8 cm]{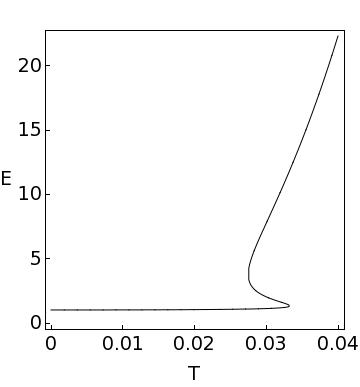}
\caption{\label{fig:energy_function_T_J}In the Kerr-AdS case, up to three black holes can exist with a given temperature $T$, for a certain range of a fixed angular momentum $J$. In this example, $J=1$ and $\ell=10$. In fact, a fourth solution exists with small values of the energy $E$, but is beyond the limit $|a|=\ell$.}
\end{figure}

We analysed, in the canonical ensemble (charges $J$ or $Q$ fixed), the cases of Kerr black hole, with or without cosmological constant, and of the Reissner-Nordstr\"om black hole. We use the definitions $C \sim \pm |T-T_c|^{-\alpha}$, for the phase $T>T_c$, and $C \sim \pm |T-T_c|^{-\alpha'}$, for the phase $T<T_c$. We will actually find that the critical temperature $T_c$ is a local maximum or minimum as a function of the energy $E$, for fixed charges $J$ or $Q$. This means that only one of the exponents $\alpha$ and $\alpha'$ is relevant. In a physical situation, on one side of the phase transition there is the stable black hole, while on the other side there is a different stable configuration (eg. spacetime with radiation \cite{York:1986it}) the specific heat of which would contribute with the other critical exponent. In the asymptotically flat cases, $T_c$ is a maximum of the temperature as a function of the energy $E$, for fixed $J$ or $Q$, which means that the relevant exponent is $\alpha'$. In the Schwarzschild-AdS case, $T_c$ is a minimum of the temperature as a function of the energy $E$, so that the relevant exponent is $\alpha$. In the Kerr-AdS case, both situations can occur depending on the parameter region, as can be seen in Fig.~(\ref{fig:energy_function_T_J}). Since, in all cases, the temperature varies quadratically with the energy around the critical point, $T-T_c \sim \pm (E-E_c)^{1/2}$, the critical exponents have the value 1/2.

In the grand-canonical ensemble, the charge is not fixed. The interesting cases are the asymptotically AdS black holes since there is no thermodynamic stability for the asymptotically flat black holes in this ensemble. In the Schwarzschild-AdS case $\Omega=0$, the specific heat is the same as in the canonical case, diverging as described above. We have also mentioned that the moment of inertia, given by $\epsilon = r_+^3/2$, is finite. When $\Omega \neq 0$, there is a sharp difference as $\epsilon$ becomes divergent over the line of critical stability. Let us recall equation (\ref{cocj}) in four dimensions,
\be
C_{\Omega} = C_J + \frac{T \eta^2}{\epsilon}.
\ee
The quantity $\eta$ diverges as $\epsilon$ and $C_{\Omega}$, while $C_J$ is finite, along the critical line. Again the divergences have a critical exponent 1/2, at fixed $\Omega$ (and the same dependence on $\Omega - \Omega_c$ for fixed $T$).

\section{\label{sec:conc}Conclusions}

In this paper, we studied the thermodynamic instabilities of several rotating black holes. We analysed their grand-canonical partition functions and found that they possess negative modes at the perturbative level. An argument connecting this pathology to the lack of thermodynamic stability is extended here to the grand-canonical ensemble.

The method we applied to look for negative modes avoids the complications due to the lack of symmetry of these solutions. Instead of addressing the full problem, we construct a non-conformal probe perturbation which lies along the stationary phase path. If the probe perturbation decreases the Euclidean action, then a negative mode exists. We also clarified the issue of using Euclidean methods for rotating spacetimes, leading to complex (quasi-Euclidean) instantons which should present no difficulty of principle.

We found our results for the negative modes to be consistent with the standard conditions of thermodynamical stability in the grand-canonical ensemble, with the exception of the Kerr-AdS black hole. In the latter case, a small parameter region near the line of critical stability presents a negative mode when the zero-th order conditions of stability are satisfied. This feature suggests that the thermodynamic limit is ill-defined near the critical line in this case, so that the location of the phase transition is corrected by quantum perturbations.

Possible future work linked to these results includes: (i) studying the partition function of a dual conformal field theory and comparing with the gravitational case; (ii) studying the full negative modes problem numerically, with special focus on the phase transition of the asymptotically AdS black holes; (iii) formulating the canonical ensemble perturbations problem, which would require fixing the angular momenta $J^i$, \emph{i.e.} writing the second-order action in terms of perturbations that would specify derivatives of the metric normal to the boundary.

\section{\label{sec:thanks}Acknowledgments}

We are grateful to Gary Gibbons, Stephen Hawking, Gustav Holzegel, Hari Kunduri and Claude Warnick for valuable discussions. RM and JES acknowledge support from the Funda\c c\~ao para a Ci\^encia e Tecnologia (FCT, Portugal) through the grants SFRH/BD/22211/2005 (RM) and SFRH/BD/22058/2005 (JES).

\bibliography{bibliography}

\begin{thebibliography}{47}
\expandafter\ifx\csname natexlab\endcsname\relax\def\natexlab#1{#1}\fi
\expandafter\ifx\csname bibnamefont\endcsname\relax
  \def\bibnamefont#1{#1}\fi
\expandafter\ifx\csname bibfnamefont\endcsname\relax
  \def\bibfnamefont#1{#1}\fi
\expandafter\ifx\csname citenamefont\endcsname\relax
  \def\citenamefont#1{#1}\fi
\expandafter\ifx\csname url\endcsname\relax
  \def\url#1{\texttt{#1}}\fi
\expandafter\ifx\csname urlprefix\endcsname\relax\def\urlprefix{URL }\fi
\providecommand{\bibinfo}[2]{#2}
\providecommand{\eprint}[2][]{\url{#2}}

\bibitem[{\citenamefont{Regge and Wheeler}(1957)}]{PhysRev.108.1063}
\bibinfo{author}{\bibfnamefont{T.}~\bibnamefont{Regge}} \bibnamefont{and}
  \bibinfo{author}{\bibfnamefont{J.~A.} \bibnamefont{Wheeler}},
  \bibinfo{journal}{Phys. Rev.} \textbf{\bibinfo{volume}{108}},
  \bibinfo{pages}{1063} (\bibinfo{year}{1957}).

\bibitem[{\citenamefont{Zerilli}(1970)}]{PhysRevLett.24.737}
\bibinfo{author}{\bibfnamefont{F.~J.} \bibnamefont{Zerilli}},
  \bibinfo{journal}{Phys. Rev. Lett.} \textbf{\bibinfo{volume}{24}},
  \bibinfo{pages}{737} (\bibinfo{year}{1970}).

\bibitem[{\citenamefont{Vishveshwara}(1970)}]{PhysRevD.1.2870}
\bibinfo{author}{\bibfnamefont{C.~V.} \bibnamefont{Vishveshwara}},
  \bibinfo{journal}{Phys. Rev. D} \textbf{\bibinfo{volume}{1}},
  \bibinfo{pages}{2870} (\bibinfo{year}{1970}).

\bibitem[{\citenamefont{Teukolsky}(1972)}]{Teukolsky:1972my}
\bibinfo{author}{\bibfnamefont{S.~A.} \bibnamefont{Teukolsky}},
  \bibinfo{journal}{Phys. Rev. Lett.} \textbf{\bibinfo{volume}{29}},
  \bibinfo{pages}{1114} (\bibinfo{year}{1972}).

\bibitem[{\citenamefont{Chandrasekhar}(1992)}]{Chandrasekhar:1985kt}
\bibinfo{author}{\bibfnamefont{S.}~\bibnamefont{Chandrasekhar}},
  \emph{\bibinfo{title}{{The mathematical theory of black holes}}}
  (\bibinfo{publisher}{Oxford, UK: Clarendon}, \bibinfo{year}{1992}).

\bibitem[{\citenamefont{Whiting}(1989)}]{Whiting:1988vc}
\bibinfo{author}{\bibfnamefont{B.~F.} \bibnamefont{Whiting}},
  \bibinfo{journal}{J. Math. Phys.} \textbf{\bibinfo{volume}{30}},
  \bibinfo{pages}{1301} (\bibinfo{year}{1989}).

\bibitem[{\citenamefont{Ishibashi and Kodama}(2003)}]{Ishibashi:2003ap}
\bibinfo{author}{\bibfnamefont{A.}~\bibnamefont{Ishibashi}} \bibnamefont{and}
  \bibinfo{author}{\bibfnamefont{H.}~\bibnamefont{Kodama}},
  \bibinfo{journal}{Prog. Theor. Phys.} \textbf{\bibinfo{volume}{110}},
  \bibinfo{pages}{901} (\bibinfo{year}{2003}), \eprint{hep-th/0305185}.

\bibitem[{\citenamefont{Hawking}(1974)}]{Hawking:1974rv}
\bibinfo{author}{\bibfnamefont{S.~W.} \bibnamefont{Hawking}},
  \bibinfo{journal}{Nature} \textbf{\bibinfo{volume}{248}}, \bibinfo{pages}{30}
  (\bibinfo{year}{1974}).

\bibitem[{\citenamefont{Hawking}(1976)}]{Hawking:1976de}
\bibinfo{author}{\bibfnamefont{S.~W.} \bibnamefont{Hawking}},
  \bibinfo{journal}{Phys. Rev.} \textbf{\bibinfo{volume}{D13}},
  \bibinfo{pages}{191} (\bibinfo{year}{1976}).

\bibitem[{\citenamefont{Gross et~al.}(1982)\citenamefont{Gross, Perry, and
  Yaffe}}]{Gross:1982cv}
\bibinfo{author}{\bibfnamefont{D.~J.} \bibnamefont{Gross}},
  \bibinfo{author}{\bibfnamefont{M.~J.} \bibnamefont{Perry}}, \bibnamefont{and}
  \bibinfo{author}{\bibfnamefont{L.~G.} \bibnamefont{Yaffe}},
  \bibinfo{journal}{Phys. Rev.} \textbf{\bibinfo{volume}{D25}},
  \bibinfo{pages}{330} (\bibinfo{year}{1982}).

\bibitem[{\citenamefont{Prestidge}(2000)}]{Prestidge:1999uq}
\bibinfo{author}{\bibfnamefont{T.}~\bibnamefont{Prestidge}},
  \bibinfo{journal}{Phys. Rev.} \textbf{\bibinfo{volume}{D61}},
  \bibinfo{pages}{084002} (\bibinfo{year}{2000}), \eprint{hep-th/9907163}.

\bibitem[{\citenamefont{York}(1972)}]{York:1972sj}
\bibinfo{author}{\bibfnamefont{J.}~\bibnamefont{York},
  \bibfnamefont{James~W.}}, \bibinfo{journal}{Phys. Rev. Lett.}
  \textbf{\bibinfo{volume}{28}}, \bibinfo{pages}{1082} (\bibinfo{year}{1972}).

\bibitem[{\citenamefont{Gibbons and Hawking}(1977)}]{Gibbons:1976ue}
\bibinfo{author}{\bibfnamefont{G.~W.} \bibnamefont{Gibbons}} \bibnamefont{and}
  \bibinfo{author}{\bibfnamefont{S.~W.} \bibnamefont{Hawking}},
  \bibinfo{journal}{Phys. Rev.} \textbf{\bibinfo{volume}{D15}},
  \bibinfo{pages}{2752} (\bibinfo{year}{1977}).

\bibitem[{\citenamefont{Hawking and Page}(1983)}]{Hawking:1982dh}
\bibinfo{author}{\bibfnamefont{S.~W.} \bibnamefont{Hawking}} \bibnamefont{and}
  \bibinfo{author}{\bibfnamefont{D.~N.} \bibnamefont{Page}},
  \bibinfo{journal}{Commun. Math. Phys.} \textbf{\bibinfo{volume}{87}},
  \bibinfo{pages}{577} (\bibinfo{year}{1983}).

\bibitem[{\citenamefont{Gibbons et~al.}(2005)\citenamefont{Gibbons, Perry, and
  Pope}}]{Gibbons:2004ai}
\bibinfo{author}{\bibfnamefont{G.~W.} \bibnamefont{Gibbons}},
  \bibinfo{author}{\bibfnamefont{M.~J.} \bibnamefont{Perry}}, \bibnamefont{and}
  \bibinfo{author}{\bibfnamefont{C.~N.} \bibnamefont{Pope}},
  \bibinfo{journal}{Class. Quant. Grav.} \textbf{\bibinfo{volume}{22}},
  \bibinfo{pages}{1503} (\bibinfo{year}{2005}), \eprint{hep-th/0408217}.

\bibitem[{\citenamefont{Balasubramanian and
  Kraus}(1999)}]{Balasubramanian:1999re}
\bibinfo{author}{\bibfnamefont{V.}~\bibnamefont{Balasubramanian}}
  \bibnamefont{and} \bibinfo{author}{\bibfnamefont{P.}~\bibnamefont{Kraus}},
  \bibinfo{journal}{Commun. Math. Phys.} \textbf{\bibinfo{volume}{208}},
  \bibinfo{pages}{413} (\bibinfo{year}{1999}), \eprint{hep-th/9902121}.

\bibitem[{\citenamefont{Kraus et~al.}(1999)\citenamefont{Kraus, Larsen, and
  Siebelink}}]{Kraus:1999di}
\bibinfo{author}{\bibfnamefont{P.}~\bibnamefont{Kraus}},
  \bibinfo{author}{\bibfnamefont{F.}~\bibnamefont{Larsen}}, \bibnamefont{and}
  \bibinfo{author}{\bibfnamefont{R.}~\bibnamefont{Siebelink}},
  \bibinfo{journal}{Nucl. Phys.} \textbf{\bibinfo{volume}{B563}},
  \bibinfo{pages}{259} (\bibinfo{year}{1999}), \eprint{hep-th/9906127}.

\bibitem[{\citenamefont{Olea}(2005)}]{Olea:2005gb}
\bibinfo{author}{\bibfnamefont{R.}~\bibnamefont{Olea}}, \bibinfo{journal}{JHEP}
  \textbf{\bibinfo{volume}{06}}, \bibinfo{pages}{023} (\bibinfo{year}{2005}),
  \eprint{hep-th/0504233}.

\bibitem[{\citenamefont{Olea}(2007)}]{Olea:2006vd}
\bibinfo{author}{\bibfnamefont{R.}~\bibnamefont{Olea}}, \bibinfo{journal}{JHEP}
  \textbf{\bibinfo{volume}{04}}, \bibinfo{pages}{073} (\bibinfo{year}{2007}),
  \eprint{hep-th/0610230}.

\bibitem[{\citenamefont{Skenderis}(2001)}]{Skenderis:2000in}
\bibinfo{author}{\bibfnamefont{K.}~\bibnamefont{Skenderis}},
  \bibinfo{journal}{Int. J. Mod. Phys.} \textbf{\bibinfo{volume}{A16}},
  \bibinfo{pages}{740} (\bibinfo{year}{2001}), \eprint{hep-th/0010138}.

\bibitem[{\citenamefont{Gibbons et~al.}(1978)\citenamefont{Gibbons, Hawking,
  and Perry}}]{Gibbons:1978ac}
\bibinfo{author}{\bibfnamefont{G.~W.} \bibnamefont{Gibbons}},
  \bibinfo{author}{\bibfnamefont{S.~W.} \bibnamefont{Hawking}},
  \bibnamefont{and} \bibinfo{author}{\bibfnamefont{M.~J.} \bibnamefont{Perry}},
  \bibinfo{journal}{Nucl. Phys.} \textbf{\bibinfo{volume}{B138}},
  \bibinfo{pages}{141} (\bibinfo{year}{1978}).

\bibitem[{\citenamefont{Gibbons and Perry}(1978)}]{Gibbons:1978ji}
\bibinfo{author}{\bibfnamefont{G.~W.} \bibnamefont{Gibbons}} \bibnamefont{and}
  \bibinfo{author}{\bibfnamefont{M.~J.} \bibnamefont{Perry}},
  \bibinfo{journal}{Nucl. Phys.} \textbf{\bibinfo{volume}{B146}},
  \bibinfo{pages}{90} (\bibinfo{year}{1978}).

\bibitem[{\citenamefont{Yano}(1970)}]{Yano}
\bibinfo{author}{\bibfnamefont{K.}~\bibnamefont{Yano}},
  \emph{\bibinfo{title}{Integral Formulas in Riemannian Geometry}}
  (\bibinfo{publisher}{Marcel Dekker, New York}, \bibinfo{year}{1970}).

\bibitem[{\citenamefont{Brown et~al.}(1991{\natexlab{a}})\citenamefont{Brown,
  Martinez, and York}}]{Brown:1990di}
\bibinfo{author}{\bibfnamefont{J.~D.} \bibnamefont{Brown}},
  \bibinfo{author}{\bibfnamefont{E.~A.} \bibnamefont{Martinez}},
  \bibnamefont{and} \bibinfo{author}{\bibfnamefont{J.}~\bibnamefont{York},
  \bibfnamefont{James~W.}}, \bibinfo{journal}{Annals N. Y. Acad. Sci.}
  \textbf{\bibinfo{volume}{631}}, \bibinfo{pages}{225}
  (\bibinfo{year}{1991}{\natexlab{a}}).

\bibitem[{\citenamefont{Brown et~al.}(1991{\natexlab{b}})\citenamefont{Brown,
  Martinez, and York}}]{Brown:1990fk}
\bibinfo{author}{\bibfnamefont{J.~D.} \bibnamefont{Brown}},
  \bibinfo{author}{\bibfnamefont{E.~A.} \bibnamefont{Martinez}},
  \bibnamefont{and} \bibinfo{author}{\bibfnamefont{J.~W.} \bibnamefont{York}},
  \bibinfo{journal}{Phys. Rev. Lett.} \textbf{\bibinfo{volume}{66}},
  \bibinfo{pages}{2281} (\bibinfo{year}{1991}{\natexlab{b}}).

\bibitem[{\citenamefont{Astefanesei and Radu}(2006)}]{Astefanesei:2005ad}
\bibinfo{author}{\bibfnamefont{D.}~\bibnamefont{Astefanesei}} \bibnamefont{and}
  \bibinfo{author}{\bibfnamefont{E.}~\bibnamefont{Radu}},
  \bibinfo{journal}{Phys. Rev.} \textbf{\bibinfo{volume}{D73}},
  \bibinfo{pages}{044014} (\bibinfo{year}{2006}), \eprint{hep-th/0509144}.

\bibitem[{\citenamefont{Schelpe}(to be published)}]{bella}
\bibinfo{author}{\bibfnamefont{A.}~\bibnamefont{Schelpe}},
  \bibinfo{journal}{PhD Thesis, University of Cambridge}  (\bibinfo{year}{to be
  published}).

\bibitem[{\citenamefont{Hawking}(1972)}]{Hawking:1972qk}
\bibinfo{author}{\bibfnamefont{S.~W.} \bibnamefont{Hawking}},
  \bibinfo{journal}{Comm. Math. Phys.} \textbf{\bibinfo{volume}{25}},
  \bibinfo{pages}{167} (\bibinfo{year}{1972}).

\bibitem[{\citenamefont{Hawking and Ellis}(1973)}]{Hawking:1973uf}
\bibinfo{author}{\bibfnamefont{S.~W.} \bibnamefont{Hawking}} \bibnamefont{and}
  \bibinfo{author}{\bibfnamefont{G.~F.~R.} \bibnamefont{Ellis}},
  \emph{\bibinfo{title}{{The Large scale structure of space-time}}}
  (\bibinfo{publisher}{{Cambridge University Press}}, \bibinfo{year}{1973}).

\bibitem[{\citenamefont{Kerr}(1963)}]{Kerr:1963ud}
\bibinfo{author}{\bibfnamefont{R.~P.} \bibnamefont{Kerr}},
  \bibinfo{journal}{Phys. Rev. Lett.} \textbf{\bibinfo{volume}{11}},
  \bibinfo{pages}{237} (\bibinfo{year}{1963}).

\bibitem[{\citenamefont{Myers and Perry}(1986)}]{Myers:1986un}
\bibinfo{author}{\bibfnamefont{R.~C.} \bibnamefont{Myers}} \bibnamefont{and}
  \bibinfo{author}{\bibfnamefont{M.~J.} \bibnamefont{Perry}},
  \bibinfo{journal}{Ann. Phys.} \textbf{\bibinfo{volume}{172}},
  \bibinfo{pages}{304} (\bibinfo{year}{1986}).

\bibitem[{\citenamefont{Emparan and Reall}(2002)}]{Emparan:2001wn}
\bibinfo{author}{\bibfnamefont{R.}~\bibnamefont{Emparan}} \bibnamefont{and}
  \bibinfo{author}{\bibfnamefont{H.~S.} \bibnamefont{Reall}},
  \bibinfo{journal}{Phys. Rev. Lett.} \textbf{\bibinfo{volume}{88}},
  \bibinfo{pages}{101101} (\bibinfo{year}{2002}), \eprint{hep-th/0110260}.

\bibitem[{\citenamefont{Chen et~al.}(2007)\citenamefont{Chen, Lu, and
  Pope}}]{Chen:2006ea}
\bibinfo{author}{\bibfnamefont{W.}~\bibnamefont{Chen}},
  \bibinfo{author}{\bibfnamefont{H.}~\bibnamefont{Lu}}, \bibnamefont{and}
  \bibinfo{author}{\bibfnamefont{C.~N.} \bibnamefont{Pope}},
  \bibinfo{journal}{Nucl. Phys.} \textbf{\bibinfo{volume}{B762}},
  \bibinfo{pages}{38} (\bibinfo{year}{2007}), \eprint{hep-th/0601002}.

\bibitem[{\citenamefont{Tangherlini}(1963)}]{Tangherlini:1963bw}
\bibinfo{author}{\bibfnamefont{F.~R.} \bibnamefont{Tangherlini}},
  \bibinfo{journal}{Nuovo Cim.} \textbf{\bibinfo{volume}{27}},
  \bibinfo{pages}{636} (\bibinfo{year}{1963}).

\bibitem[{\citenamefont{Kleihaus et~al.}(2007)\citenamefont{Kleihaus, Kunz, and
  Radu}}]{Kleihaus:2007dg}
\bibinfo{author}{\bibfnamefont{B.}~\bibnamefont{Kleihaus}},
  \bibinfo{author}{\bibfnamefont{J.}~\bibnamefont{Kunz}}, \bibnamefont{and}
  \bibinfo{author}{\bibfnamefont{E.}~\bibnamefont{Radu}},
  \bibinfo{journal}{JHEP} \textbf{\bibinfo{volume}{05}}, \bibinfo{pages}{058}
  (\bibinfo{year}{2007}), \eprint{hep-th/0702053}.

\bibitem[{\citenamefont{Reall}(2001)}]{Reall:2001ag}
\bibinfo{author}{\bibfnamefont{H.~S.} \bibnamefont{Reall}},
  \bibinfo{journal}{Phys. Rev.} \textbf{\bibinfo{volume}{D64}},
  \bibinfo{pages}{044005} (\bibinfo{year}{2001}), \eprint{hep-th/0104071}.

\bibitem[{\citenamefont{Chen et~al.}(2006)\citenamefont{Chen, Lu, and
  Pope}}]{Chen:2006xh}
\bibinfo{author}{\bibfnamefont{W.}~\bibnamefont{Chen}},
  \bibinfo{author}{\bibfnamefont{H.}~\bibnamefont{Lu}}, \bibnamefont{and}
  \bibinfo{author}{\bibfnamefont{C.~N.} \bibnamefont{Pope}},
  \bibinfo{journal}{Class. Quant. Grav.} \textbf{\bibinfo{volume}{23}},
  \bibinfo{pages}{5323} (\bibinfo{year}{2006}), \eprint{hep-th/0604125}.

\bibitem[{\citenamefont{Yang}(1999)}]{312597}
\bibinfo{author}{\bibfnamefont{L.}~\bibnamefont{Yang}}, \bibinfo{journal}{J.
  Symb. Comput.} \textbf{\bibinfo{volume}{28}}, \bibinfo{pages}{225}
  (\bibinfo{year}{1999}), ISSN \bibinfo{issn}{0747-7171}.

\bibitem[{\citenamefont{Hawking et~al.}(1999)\citenamefont{Hawking, Hunter, and
  Taylor-Robinson}}]{Hawking:1998kw}
\bibinfo{author}{\bibfnamefont{S.~W.} \bibnamefont{Hawking}},
  \bibinfo{author}{\bibfnamefont{C.~J.} \bibnamefont{Hunter}},
  \bibnamefont{and} \bibinfo{author}{\bibfnamefont{M.~M.}
  \bibnamefont{Taylor-Robinson}}, \bibinfo{journal}{Phys. Rev.}
  \textbf{\bibinfo{volume}{D59}}, \bibinfo{pages}{064005}
  (\bibinfo{year}{1999}), \eprint{hep-th/9811056}.

\bibitem[{\citenamefont{York}(1986)}]{York:1986it}
\bibinfo{author}{\bibfnamefont{J.~W.} \bibnamefont{York}},
  \bibinfo{journal}{Phys. Rev.} \textbf{\bibinfo{volume}{D33}},
  \bibinfo{pages}{2092} (\bibinfo{year}{1986}).

\bibitem[{\citenamefont{Hawking and Reall}(1999)}]{Hawking:1999dp}
\bibinfo{author}{\bibfnamefont{S.~W.} \bibnamefont{Hawking}} \bibnamefont{and}
  \bibinfo{author}{\bibfnamefont{H.~S.} \bibnamefont{Reall}},
  \bibinfo{journal}{Phys. Rev.} \textbf{\bibinfo{volume}{D61}},
  \bibinfo{pages}{024014} (\bibinfo{year}{1999}), \eprint{hep-th/9908109}.

\bibitem[{\citenamefont{Landau and Lifschitz}(1958)}]{Landau}
\bibinfo{author}{\bibfnamefont{L.~D.} \bibnamefont{Landau}} \bibnamefont{and}
  \bibinfo{author}{\bibfnamefont{E.~M.} \bibnamefont{Lifschitz}},
  \emph{\bibinfo{title}{Statistical Physics, Part 1}}
  (\bibinfo{publisher}{Pergamon Press, Oxford}, \bibinfo{year}{1958}).

\bibitem[{\citenamefont{Weinhold}(1975)}]{Weinhold}
\bibinfo{author}{\bibfnamefont{F.}~\bibnamefont{Weinhold}},
  \bibinfo{journal}{J. Chem. Phys.} \textbf{\bibinfo{volume}{63}},
  \bibinfo{pages}{2479, 2484, 2488, 2486} (\bibinfo{year}{1975}).

\bibitem[{\citenamefont{Ruppeiner}(1995)}]{Ruppeiner}
\bibinfo{author}{\bibfnamefont{G.}~\bibnamefont{Ruppeiner}},
  \bibinfo{journal}{Rev. Mod. Phys.} \textbf{\bibinfo{volume}{67}},
  \bibinfo{pages}{605} (\bibinfo{year}{1995}).

\bibitem[{\citenamefont{Whiting and York}(1988)}]{PhysRevLett.61.1336}
\bibinfo{author}{\bibfnamefont{B.~F.} \bibnamefont{Whiting}} \bibnamefont{and}
  \bibinfo{author}{\bibfnamefont{J.~W.} \bibnamefont{York}},
  \bibinfo{journal}{Phys. Rev. Lett.} \textbf{\bibinfo{volume}{61}},
  \bibinfo{pages}{1336} (\bibinfo{year}{1988}).

\bibitem[{\citenamefont{Stanley}(1971)}]{Stanley}
\bibinfo{author}{\bibfnamefont{H.~E.} \bibnamefont{Stanley}},
  \emph{\bibinfo{title}{Introduction to Phase Transitions and Critical
  Phenomena}} (\bibinfo{publisher}{Oxford University Press},
  \bibinfo{year}{1971}).

\bibitem[{\citenamefont{Monteiro and Santos}(2009)}]{monteiro:064006}
\bibinfo{author}{\bibfnamefont{R.}~\bibnamefont{Monteiro}} \bibnamefont{and}
  \bibinfo{author}{\bibfnamefont{J.~E.} \bibnamefont{Santos}},
  \bibinfo{journal}{Phys. Rev.} \textbf{\bibinfo{volume}{D79}},
  \bibinfo{pages}{064006} (\bibinfo{year}{2009}), \eprint{0812.1767}.

\end{thebibliography}

\end{document}